\documentclass[manuscript,screen,review=false,natbib=true]{acmart}

\usepackage{siunitx}
\usepackage{multirow}
\usepackage[dvipsnames, HTML]{xcolor}
\usepackage{hyperref}
\usepackage{comment}
\usepackage[abbreviations]{glossaries-extra}
\usepackage{subcaption}

\newabbreviation{abft}{ABFT}{Algorithm-Based Fault Tolerance}
\newabbreviation{ai}{AI}{Artificial Intelligence}
\newabbreviation{bist}{BIST}{Built-In-Self-Test}
\newabbreviation{cnn}{CNN}{Convolutional Neural Network}
\newabbreviation{cots}{COTS}{Commercial-off-the-shelf}
\newabbreviation{dma}{DMA}{Direct Memory Access}
\newabbreviation{dmr}{DMR}{Dual Modular Redundancy}
\newabbreviation{dnn}{DNN}{Deep Neural Network}
\newabbreviation{dppu}{DPPU}{Dot-Production Processing Unit}
\newabbreviation{dsp}{DSP}{Digital Signal Processor}
\newabbreviation{ecc}{ECC}{Error Correction Code}
\newabbreviation{fdsoi}{FD-SOI}{Fully Depleted Silicon On Insulator}
\newabbreviation{fpu}{FPU}{Floating-Point Unit}
\newabbreviation{fsm}{FSM}{Finite State Machine}
\newabbreviation{gemm}{GEMM}{General Matrix Multiply}
\newabbreviation{geo}{GEO}{Geostationary Orbit}
\newabbreviation{gnc}{GNC}{Guidance, Navigation, and Control}
\newabbreviation{hci}{HCI}{Heterogeneous Cluster Interconnect}
\newabbreviation{hls}{HLS}{High Level Synthesis}
\newabbreviation{hmr}{HMR}{Hybrid Modular Redundancy}
\newabbreviation{hwpe}{HWPE}{Hardware Processing Engine}
\newabbreviation{imc}{IMC}{In-Memory Computing}
\newabbreviation{leo}{LEO}{Low Earth Orbit}
\newabbreviation{mac}{MAC}{Multiply-Accumulate}
\newabbreviation{mbu}{MBU}{Multiple Bit Upset}
\newabbreviation{nn}{NN}{Neural Network}
\newabbreviation{npu}{NPU}{Neural Processing Unit}
\newabbreviation{oec}{OEC}{Orbital Edge Computing}
\newabbreviation{pe}{PE}{Processing Element}
\newabbreviation{ppa}{PPA}{Power, Performance and Area}
\newabbreviation{pulp}{PULP}{Parallel Ultra-Low-Power}
\newabbreviation{rcm}{RCM}{Re-Computing Module}
\newabbreviation{rhbd}{RHBD}{Radiation-Hardening-By-Design}
\newabbreviation{rhbp}{RHBP}{Radiation-Hardening-By-Process}
\newabbreviation{ru}{RU}{Re-computing Unit}
\newabbreviation{sbu}{SBU}{Single Bit Upset}
\newabbreviation{sdc}{SDC}{Silent Data Corruption}
\newabbreviation{secded}{SEC-DED}{Single Error Correction, Double Error Detection}
\newabbreviation{sed}{SED}{Symptom-Based Error Detector}
\newabbreviation{see}{SEE}{Single Event Effect}
\newabbreviation{set}{SET}{Single Event Transient}
\newabbreviation{seu}{SEU}{Single Event Upset}
\newabbreviation{slh}{SLH}{Selective Latch Hardening}
\newabbreviation{soa}{SoA}{State-of-the-Art}
\newabbreviation{taso}{TASO}{Trusted Autonomous Satellite Operations}
\newabbreviation{tcdm}{TCDM}{Tightly Coupled Data Memory}
\newabbreviation{tmr}{TMR}{Triple Modular Redundancy}
\newabbreviation{zbma}{ZBMA}{Zero-Biased Multiple-Node-Upset-Aware}

\glsdisablehyper

\acmJournal{JETC}

\AtBeginDocument{%
  }

\setcopyright{acmlicensed}
\copyrightyear{2026}
\acmYear{2026}
\acmDOI{XXXXXXX.XXXXXXX}
\acmISBN{978-1-4503-XXXX-X/2018/06}

\begin{document}

\title{Safe-NEureka: a Hybrid Modular Redundant DNN Accelerator for On-board Satellite AI Processing}

\titlenote{This article is an extended version of a paper entitled "HMR-NEureka: Hybrid Modular Redundancy DNN Acceleration in Heterogeneous RISC-V SoCs," presented at the  2025 IEEE Computer Society Annual Symposium on VLSI (ISVLSI).}

\author{Riccardo Tedeschi}
\orcid{0009-0007-4483-9261}
\email{riccardo.tedeschi6@unibo.it}
\thanks{This work was partially supported by the Chips Joint Undertaking (CHIPS-JU) and its member Italy through the TRISTAN project (g.a. 101095947) and the ISOLDE project (g.a. 101112274).}
\authornote{Both authors contributed equally to this work.}
\affiliation{%
  \institution{University of Bologna}
  \department{Department of Electrical, Electronic and Information Engineering}
  \streetaddress{Viale del Risorgimento, 2}
  \postcode{40136}
  \city{Bologna}
  \country{Italy}
}

\author{Luigi Ghionda}
\orcid{0009-0000-5748-1085}
\authornotemark[2]
\email{luigi.ghionda@chips.it}
\affiliation{%
  \institution{Fondazione Chips-IT}
  \department{Digital Design and Open Hardware}
  \city{Bologna}
  \country{Italy}
}

\author{Alessandro Nadalini}
\orcid{0009-0007-3574-7576}
\email{alessandro.nadalini3@unibo.it}
\affiliation{%
  \institution{University of Bologna}
  \department{Department of Electrical, Electronic and Information Engineering}
  \streetaddress{Viale del Risorgimento, 2}
  \postcode{40136}
  \city{Bologna}
  \country{Italy}
}

\author{Yvan Tortorella}
\orcid{0000-0001-8248-5731}
\email{yvan.tortorella@chips.it}
\affiliation{%
  \institution{Fondazione Chips-IT}
  \department{Digital Design and Open Hardware}
  \city{Bologna}
  \country{Italy}
}

\author{Arpan Suravi Prasad}
\orcid{0009-0009-6031-6668}
\email{prasadar@iis.ee.ethz.ch}
\affiliation{%
  \institution{ETH Zürich}
  \department{Integrated Systems Laboratory}
  \streetaddress{Gloriastrasse 35}
  \postcode{8092}
  \city{Zürich}
  \country{Switzerland}
}

\author{Luca Benini}
\orcid{0000-0001-8068-3806}
\email{luca.benini@unibo.it}
\affiliation{%
  \institution{University of Bologna}
  \department{Department of Electrical, Electronic and Information Engineering}
  \streetaddress{Viale del Risorgimento, 2}
  \postcode{40136}
  \city{Bologna}
  \country{Italy}
}
\affiliation{%
  \institution{ETH Zürich}
  \department{Integrated Systems Laboratory}
  \streetaddress{Gloriastrasse 35}
  \postcode{8092}
  \city{Zürich}
  \country{Switzerland}
}

\author{Davide Rossi}
\orcid{0000-0002-0651-5393}
\email{davide.rossi@unibo.it}
\affiliation{%
  \institution{University of Bologna}
  \department{Department of Electrical, Electronic and Information Engineering}
  \streetaddress{Viale del Risorgimento, 2}
  \postcode{40136}
  \city{Bologna}
  \country{Italy}
}
\affiliation{%
  \institution{Fondazione Chips-IT}
  \department{Digital Design and Open Hardware}
  \city{Bologna}
  \country{Italy}
}

\author{Francesco Conti}
\orcid{0000-0002-7924-933X}
\email{f.conti@unibo.it}
\affiliation{%
  \institution{University of Bologna}
  \department{Department of Electrical, Electronic and Information Engineering}
  \streetaddress{Viale del Risorgimento, 2}
  \postcode{40136}
  \city{Bologna}
  \country{Italy}
}

\renewcommand{\shortauthors}{Tedeschi, Ghionda et al.}

\begin{abstract}
The deployment of Low Earth Orbit (LEO) constellations is revolutionizing the space sector, with on-board Artificial Intelligence (AI) becoming a key element for the next generation of satellites. AI acceleration is essential in safety-critical functions, such as orbital maneuvers and collision avoidance where errors cannot be tolerated, as well as in performance-critical and energy-constrained analysis and classification of data from on-board high-bandwidth sensors (e.g., for Earth observation) which can tolerate occasional errors. Hence, AI accelerators for satellites require strong protection against radiation-induced errors when used for safety-critical tasks, coupled with high throughput and efficiency when running performance-critical tasks. This paper presents \textit{Safe-NEureka}, a Hybrid Modular Redundant accelerator designed for Deep Neural Network (DNN) inference within heterogeneous RISC-V System-on-Chips (SoCs). Safe-NEureka features two operational modes: a \textit{redundancy} mode leveraging Dual Modular Redundancy (DMR) with low-overhead hardware-based recovery, and a \textit{performance} mode that repurposes redundant datapaths to maximize throughput for performance-critical tasks. Additionally, to harden the entire accelerator, Error Correction Codes (ECCs) protect its memory interface, while the controller—compact yet critical for fault tolerance—relies on Triple Modular Redundancy (TMR). Implementation results in GlobalFoundries 12nm technology demonstrate a 96\% reduction in faulty executions in \textit{redundancy} mode, with a manageable 15\% area overhead compared to a non-fault-tolerant baseline. In \textit{performance} mode, the architecture achieves near-baseline speeds on a 3$\times$3 dense convolution layer, with only a 5\% throughput penalty and an 11\% drop in energy efficiency. This contrasts with the \textit{redundancy} mode, where throughput reduces by 48\% and efficiency by 53\%. This flexibility allows the system to incur high overheads only when processing critical workloads. These findings establish Safe-NEureka as a flexible and efficient AI accelerator for space applications.
\end{abstract}



\received{20 February 2007}
\received[revised]{12 March 2009}
\received[accepted]{5 June 2009}

\maketitle

\section{Introduction}

The deployment of \gls{leo} nanosatellite constellations has revolutionized the space industry, enabling a wide array of applications ranging from global telecommunications connectivity to scientific research and Earth observation \cite{sweetingModernSmallSatellitesChanging2018}. Within this expanding ecosystem, the growing reliance on \gls{ai}, particularly for data-intensive missions like remote sensing, has exposed the limitations of traditional architectures where satellites act solely as data relays to ground stations \cite{denbyOrbitalEdgeComputing2020}. As constellation sizes increase, the aggregate volume of raw sensor data vastly exceeds available downlink bandwidth and ground station availability. This bottleneck necessitates \gls{oec} to process data onboard and downlink only actionable insights \cite{denbyOrbitalEdgeComputing2020, furanoUseArtificialIntelligence2020}.

However, the severe size, weight, and power constraints governing these systems \cite{denbyOrbitalEdgeComputing2020} compel a single onboard computer to satisfy a strict set of conflicting requirements. On one hand, the congested orbital environment necessitates autonomous \gls{gnc} capabilities—such as collision avoidance—where safety is paramount \cite{kothariFinalFrontierDeep2020, thangavelArtificialIntelligenceTrusted2024}. In these scenarios, a single bit-flip could result in a false negative detection or an erroneous maneuver, threatening the asset's survival \cite{rechArtificialNeuralNetworks2024}. This workload demands maximum reliability and hardware redundancy to mask faults, even at the cost of performance. Conversely, modern payloads like hyperspectral imagers generate raw data volumes that require maximum parallel throughput to filter non-valuable data in real-time \cite{furanoAISpaceApplications2020}, such as cloud detection algorithms during Earth observation \cite{jeppesenCloudDetectionAlgorithm2019}. In this context, occasional transient output errors are tolerable in exchange for the processing speed required to overcome the downlink bottleneck \cite{furanoUseArtificialIntelligence2020}.

Satisfying these opposing computational demands is complicated by the hostile space environment. Spacecraft computers must guarantee operational resilience against extreme radiation phenomena that drastically exceed terrestrial baselines ~\cite{rechArtificialNeuralNetworks2024}, including destructive events such as solar flares \cite{hubertImpactSolarFlares2010}. However, the continuous downscaling of physical device geometries renders integrated circuits increasingly susceptible to radiation-induced \glspl{see} ~\cite{tangSoftErrorReliability2014}. This vulnerability stems from two opposing trends: while lower supply voltages reduce the critical charge needed to upset a node, shrinking dimensions simultaneously decrease the sensitive cross-section~\cite{kobayashiScalingTrendsDigital2021}. Consequently, even if the upset rate of individual cells stabilizes, the massive integration of state-holding cells and macros results in a net increase in the overall soft error rate at the architectural level~\cite{liProcessorDesignSoft2016}. To ensure reliable operation despite these errors, space-oriented devices must implement either \gls{rhbp} (relying on specialized manufacturing to enhance radiation resistance) or \gls{rhbd}; in this latter case, designers need to introduce redundancy, either at the hardware or information level, such as \gls{dmr}, \gls{tmr} and \glspl{ecc}.
While these redundancy techniques bolster fault tolerance, they impose substantial design complexity alongside area and performance overheads~\cite{europeancooperationforspacestandardizationecssECSSEHB2040AEngineeringTechniques2023}. For instance, traditional spatial replication imposes area penalties of at least 2$\times$ and 3$\times$ for \gls{dmr} and \gls{tmr}, respectively; these overheads are difficult to accommodate within the strict power and area budgets of nanosatellites.

Implementing static redundancy on a single chip is therefore suboptimal; it is inefficient for payload processing due to large area and power overheads, yet removing it makes the system unsafe for \gls{gnc}. Consequently, implementing a unified architecture capable of dynamically reconfiguring its internal resources between these opposing modes emerges as a compelling design strategy. \gls{hmr}~\cite{rogenmoserOnDemandRedundancyGrouping2022,rogenmoserHybridModularRedundancy2023} offers a flexible solution for these systems by enabling multiple operating modes that balance performance and error resilience at the microarchitectural level. While \citet{rogenmoserHybridModularRedundancy2023} demonstrated the effectiveness of this concept via hybrid configurable dual- or triple-modular redundancy on a cluster of homogeneous RISC-V cores, their work targeted general-purpose systems without provisions for AI acceleration. However, the inherent flexibility of such clustered cores can be efficiently coupled with domain-specific AI accelerators.

To extend end-to-end fault tolerance to AI workloads while leveraging the performance advantages of heterogeneous computing, we present \textit{Safe-NEureka}, extending the previous work of \citet{ghiondaHMRNEurekaHybridModular2025}. \textit{Safe-NEureka} is based on NEureka~\cite{prasadSpecializationMeetsFlexibility2023}, a high-performance open-source hardware accelerator for \glspl{dnn} meant for integration in multi-core RISC-V clusters.
The design of Safe-NEureka partitions a 4$\times$4 datapath into two 4$\times$2 sub-units that can be dynamically configured in two modes: \textit{redundancy} mode, leveraging \gls{dmr} execution with temporal diversity between the sub-units and hardware-based recovery for fault tolerance, and \textit{performance} mode, where the sub-units operate in parallel to enhance performance in non-critical scenarios.
Unlike pure, hardcoded spatial redundancy, this runtime reconfigurability ensures that the performance and area costs of replication are not fixed overheads, but can be dynamically tuned to the specific needs of the workload. In both modes, the accelerator controller logic is protected via TMR due to its criticality in ensuring correct operation, while occupying a small fraction of the total area. We demonstrate Safe-NEureka integrated with a multi-core RISC-V cluster based on the \gls{pulp} template \cite{nadalini3TOPSRISCV2023}, augmented for error resilience by introducing \glspl{ecc} in all memories and interconnects and featuring \gls{hmr} protected cores.
To summarize, the contributions of this work are as follows:
\begin{enumerate}
\item The design of a reconfigurable \gls{hmr} datapath for a neural engine that exploits inherent accelerator parallelism to dynamically switch between a high-performance parallel mode and a safety-critical redundant mode to enhance resilience against \glspl{see}.
\item The integration of temporal diversity between datapath replicas to mitigate common-mode faults, combined with the selective application of \gls{tmr} to the controller logic.
\item A low-overhead, bounded-latency hardware-assisted recovery mechanism to ensure continuous safe operation upon error detection.
\item A fully synthesizable RTL implementation demonstrating the accelerator integrated into a multi-core RISC-V cluster with \gls{ecc}-protected interconnects and \gls{hmr}-hardened RISC-V cores for end-to-end resilience.
\end{enumerate}

We validate the proposed architecture through a complete, tapeout-ready implementation in GlobalFoundries 12nm technology. We comprehensively evaluate Safe-NEureka in both operating modes, assessing performance, energy efficiency, and latency, as well as fault tolerance. Validation via gate-level fault injection indicates a 96\% reduction in faulty executions in \textit{redundancy} mode. Furthermore, we quantify the operational costs: while the reliable mode reduces throughput by 48\% and energy efficiency by 53\%, the \textit{performance} mode operates with a reduced 5\% throughput penalty and an 11\% energy efficiency drop on a representative 3$\times$3 dense convolutional layer. These capabilities are achieved with a total area overhead of 15\% compared to the non-fault-tolerant baseline. To the best of our knowledge, this represents the first space-oriented edge AI neural engine with \gls{hmr} support. We release the full RTL design of Safe-NEureka as open-source\footnote{https://github.com/pulp-platform/neureka/tree/safe-neureka}.
\section{Background}
\subsection{Radiation induced single event effects in space}
Ionizing radiations in the space environment induce multiple reliability issues in electronic components, ranging from transient data corruption to permanent damage \cite{rechArtificialNeuralNetworks2024}. The primary cause of failure are \glspl{see}, specifically \glspl{set} and \glspl{seu}, which are transient faults that affect combinational and sequential elements, respectively, leading to incorrect logic states \cite{gaillardSingleEventEffects2011}. The underlying mechanism leading to a soft error is the strike of a radiation particle and the subsequent generation of charge carriers due to ionization. These carriers accumulate in the reversed junctions of transistors, causing unwanted voltage spikes \cite{baumannRadiationinducedSoftErrors2005}. Given the highly parallel architecture of accelerators, a single fault in a critical unit (such as a scheduler) or a shared resource can propagate to multiple independent \glspl{pe}. This propagation risks corrupting the final \gls{dnn} output, compromising inference accuracy, or inducing unrecoverable states and deadlocks that necessitate a soft reset~\cite{rechArtificialNeuralNetworks2024, ibrahimMachineLearningTechniques2020}.

As technology scales down to smaller device features to accommodate the high performance required by accelerators, two trends affect device vulnerability. On one hand, the critical charge required to alter a node's logic state decreases with supply voltage. On the other hand, smaller feature sizes reduce the vulnerable cross-section of individual sequential elements \cite{kobayashiScalingTrendsDigital2021, tangSoftErrorReliability2014}. However, for \gls{dnn} accelerators, the massive integration of processing elements and state bits offsets this benefit, driving an overall rise in soft error rates despite the reduced per-bit vulnerability \cite{liProcessorDesignSoft2016, ibrahimMachineLearningTechniques2020}. Furthermore, the high clock frequencies and low supply voltages used to maximize efficiency in these devices exacerbate vulnerability to \glspl{set}, and technology scaling increases the occurrence of \glspl{mbu} compared to \glspl{sbu} \cite{tangSoftErrorReliability2014}. The \gls{see} rate in space varies significantly depending on the orbit and environmental conditions; for instance, a commercial 28 nm \gls{fdsoi} SRAM in \gls{geo} experiences an upset rate of approximately $4.66 \times 10^{-9}$ upsets/bit/day during solar minimum \cite{malherbeOnOrbitUpsetRate2017}. In contrast, rates in \gls{leo} are typically lower, estimated around $5 \times 10^{-10}$ upsets/bit/day, whereas extreme solar events in \gls{geo} can temporarily spike the rate to as high as $10^{-2}$ upsets/bit/day \cite{dimascioOpensourceIPCores2021}. To cope with soft errors, multiple techniques acting at different abstraction levels have been proposed to mitigate faults.

\subsection{Reliability techniques for accelerators}

\subsubsection{Software and algorithmic approaches}
These techniques rely on the intrinsic fault tolerance of \glspl{dnn}, modifying the model, the training process, or the software execution flow without changing the underlying hardware architecture.

Targeting the training phase, \citet{leeFaultToleranceAnalysis2014} introduce a weight dropout technique, where a percentage of weight connections are randomly zeroed out during training. This approach increases the network's tolerance to static interconnection and processing unit faults.
However, training improvements cannot fully mitigate faults affecting significant bits. Although \glspl{dnn} are intrinsically robust to minor perturbations \cite{mittalSurveyModelingImproving2020}, errors in high-order bits can still cause drastic value deviations and numerical instability. To mitigate this, \citet{hoangFTClipActResilienceAnalysis2020} propose \textit{FT-ClipAct}, replacing standard unbounded activation functions with clipped variants. These constrain layer outputs to a statistically defined range, effectively masking extreme values caused by faults. Similarly, \citet{chenLowcostFaultCorrector2025} present \textit{Ranger} as a low-cost fault corrector by utilizing offline profiling to identify typical computation ranges. It applies runtime restrictions to selected layers (such as activation and pooling) to proactively truncate out-of-bound values.

Taking a detection-oriented approach, \citet{liUnderstandingErrorPropagation2017} evaluate the fault-tolerance of DNN models on a dedicated ASIC accelerator. They propose a Symptom-Based Error Detector (SED) that identifies faults by monitoring activation values for deviations from valid ranges defined during profiling. This software-based technique mitigates \gls{sdc} by asynchronously checking global buffer values against these thresholds.

Finally, observing that \gls{gemm} kernels account for approximately 67\% of the processing time in the Darknet \gls{dnn}, \citet{fernandesdossantosEvaluationMitigationSoftErrors2017} implement an \gls{abft} strategy to improve reliability. This is achieved by substituting the final row and column of the input matrices with checksum vectors. This design preserves the original matrix dimensions to avoid performance overhead, accepting a minor reduction in accuracy. Consequently, the resulting product matrix inherently includes checksums, which serve as a reference for comparison against recomputed values to detect and correct radiation-induced errors.

\subsubsection{Sensitivity-Based Mapping \& Scheduling}
These techniques employ software-level analysis to optimize hardware utilization, primarily by distinguishing between critical and non-critical data.

A weight sensitivity-based design for energy-efficient DNN accelerators is proposed by \citet{choiSensitivityBasedError2019} to address timing errors caused by aggressive voltage scaling. This method leverages Taylor expansion-based sensitivity analysis to identify critical weights and filters. These critical components are assigned to robust \gls{mac} units while less sensitive computations are allocated to less reliable ones, thereby achieving fault tolerance without requiring additional redundancy.

Similarly, \citet{schornAccurateNeuronResilience2018} introduce a fine-grained analysis method that uses layer-wise relevance propagation and Taylor decomposition to predict the error resilience of individual neurons based on their contribution to the classification outcome. This resilience ranking guides the selective mapping of critical neurons (those with low resilience) to protected \glspl{pe}, while assigning less critical computations to potentially unreliable or approximate hardware units.

Finally, \citet{weiHApFTHybridApproximate2024} propose a hybrid approximate fault tolerance framework (\textit{HAp-FT}) that combines proactive risk transfer with group-level error detection. Specifically, they define a weight vulnerability metric to identify high-risk parameters, which are then distributed among neighboring \glspl{pe}. Additionally, similar filters are remapped to adjacent columns to perform online consistency checking between them, flagging output mismatches that exceed a defined threshold.

\subsubsection{Hardware \& Architecture Level}
These techniques necessitate modifications to the physical circuit design, memory cells, or architectural components of the accelerator, yet they generally remain transparent to the software stack. \gls{rhbp} leverages processes inherently resistant to radiation effects, such as \gls{fdsoi}, epitaxial layers, triple-well, and buried layers~\cite{europeancooperationforspacestandardizationecssECSSEHB2040AEngineeringTechniques2023}. However, the limited production volumes characteristic of niche markets like defense and aerospace make it difficult to justify the cost of ad-hoc \gls{rhbp} in scaled technology nodes~\cite{ginosarSurveyProcessorsSpace2012}, which are essential for high-performance designs such as accelerators. Consequently, \gls{rhbd} is often preferred, as it explicitly addresses fault tolerance through design choices at the circuit, architectural, and system levels~\cite{ibrahimMachineLearningTechniques2020}.

To address vulnerabilities in storage elements, several circuit-level modifications have been proposed. \citet{liUnderstandingErrorPropagation2017} introduce \gls{slh} as a technique that optimizes standard latch hardening by applying redundant circuitry only to the most sensitive storage elements identified through fault injection profiling. Similarly, \citet{azizimazreahToleratingSoftErrors2018} present a specialized SRAM cell known as \textit{\gls{zbma}} to mitigate multiple-node upsets. The \gls{zbma} cell exploits the inherent bias toward zero in \gls{dnn} data, including feature maps and weights, to enhance resilience.

Addressing the increasing prevalence of \glspl{mbu} in scaled technologies, \citet{parkPoPECCRobustFlexible2025} propose \textit{PoP-ECC} (Parities of Parities), a two-tier error correction scheme designed for on-chip \gls{dnn} memories. Unlike traditional single-tier codes that struggle with multi-bit errors, PoP-ECC constructs a secondary layer of redundancy derived from virtual parities of the weight matrices to specifically target complex upset patterns. To offset the storage overhead of this protection, the authors introduce a co-design strategy, \textit{Q+PoP}, which applies channel-wise quantization to reduce data precision, thereby reclaiming memory space to accommodate the additional \gls{ecc} bits without expanding the physical memory footprint.

At the processing element level, \citet{liSoftErrorMitigation2020} identify persistent SEUs in \glspl{pe}, particularly those causing large positive perturbations, as the primary driver of accuracy loss in FPGA-based CNN accelerators. To counter this, they introduce a detection scheme that leverages inherent idle cycles to locate faulty \glspl{pe} via predefined computations without incurring performance penalties. Detected faults are mitigated using an error masking technique that zeroes out the faulty \gls{pe}'s output, effectively halting error propagation. In a similar fashion, \citet{salamiResilienceRTLNN2018} investigate RTL-level vulnerability in fully connected \gls{nn} accelerators. Their proposed technique relies on the sparsity of \gls{nn} data to mask stuck-at bits through a tiered approach: \textit{Sign-Bit Masking}, which recovers a corrupted sign bit using the highly correlated MSB; \textit{Bit Masking}, which overwrites faulty bits with the sign bit to maintain small values; and \textit{Word Masking}, which resets the entire register to zero if the MSB is corrupted.

Finally, several works propose architectural redundancy to mitigate faults. In the Aphelios \gls{npu}, \citet{sunApheliosSelectiveLockstep2025} selectively duplicate critical logic (e.g., control paths) and utilize a redundant, reduced-size systolic array to recompute only the most significant output neurons, which are identified via hardware-based dynamic sorting. To mitigate permanent faults \cite{mittalSurveyModelingImproving2020} without disrupting the systolic array's synchronization, several works employ a decoupled re-compute strategy. \citet{zhaoFSAEfficientFaulttolerant2022} introduce \textit{FSA}, which isolates faulty \glspl{pe} identified via \gls{bist} by zeroing their partial sums to preserve the main datapath's dataflow. Simultaneously, a parallel \gls{rcm} fetches the corresponding inputs from on-chip buffers, a \gls{ru} calculates the missing products, and injects the correct partial sums back into the array to fill the zeroed slots. Adopting a similar decoupled approach, \citet{liuHyCAHybridComputing2022} propose \textit{HyCA}, where the main array executes without stalling, potentially generating incorrect results, while a set of global spare \glspl{dppu} runs in parallel. Leveraging a \gls{bist}-generated table to track faulty \glspl{pe}, these units retrieve operands from dedicated register files to recompute the operations of defective \glspl{pe}, correcting the final output by overwriting the erroneous results in memory.

To implement redundancy at layer-level granularity, \citet{libanoSelectiveHardeningNeural2019} optimize FPGA reliability by distinguishing between tolerable errors, which affect output values without altering classification, and critical errors that result in incorrect predictions. Through exhaustive fault injection, each network layer is profiled to identify those most susceptible to generating these critical failures, and \gls{tmr} is then applied exclusively to these high-sensitivity layers.

Scaling this redundancy up to the task level, \citet{syedFPGAImplementationFaultTolerant2024} propose a shared-layers \gls{cnn} accelerator framework that leverages a branched model to perform multi-task learning. By deploying redundant hardware instances, the system supports three reconfigurable operating modes: a \textit{Fault-Tolerant} (FT) mode, which utilizes modular redundancy to execute identical tasks across replicas for high reliability; a \textit{High-Performance} (HP) mode, which enables parallel task execution across different replicas to maximize throughput; and a \textit{De-Stress} (DS) mode, which employs partial reconfiguration to erase idle replicas and alternates the active instance to mitigate circuit aging.

Targeting the front-end electronics of the CMS High-Granularity Calorimeter (HGCAL) at CERN, \citet{guglielmoReconfigurableNeuralNetwork2021} present the design of \textit{ECON-T}, an ASIC \gls{nn} accelerator for lossy data compression that incorporates distinct fault-tolerance strategies at different abstraction levels. Implemented in 65-nm LP-CMOS using normal threshold standard cells, the design explicitly avoids minimum-sized cells to counteract radiation-induced timing degradation. \glspl{see} are mitigated via a hybrid \gls{tmr} scheme: high-speed data paths rely on register-only triplication due to high refresh rates, while static configuration blocks employ full module triplication with auto-correction to prevent error accumulation in the weights. Furthermore, the architecture avoids SRAM in favor of triplicated registers.

\subsection{Fault tolerance in emerging acceleration paradigms}
Beyond standard digital accelerators, \gls{imc} architectures maximize energy efficiency but encounter distinct reliability issues arising from the specific memory technologies employed.
Emerging technology-based IMCs, such as those using ReRAM, are hindered by relatively immature manufacturing processes that lead to high rates of permanent stuck-at faults and analog non-idealities like conductance variations \cite{malhotraInvitedPaperFault2024}.
Mitigation strategies for these defects often rely on model adaptation, such as on-line fault detection coupled with fault-tolerant training to handle both hard and soft faults \cite{xiaFaulttolerantTrainingOnline2017}.
Alternatively, hardware-centric solutions optimize resilience by utilizing redundancy strategies that group crossbars and map weights based on layer sensitivity \cite{guoATTFaultTolerantReRAM2020}.
Conversely, Digital Computing-in-Memory (DCIM) architectures utilizing mature SRAM technology compromise the inherent robustness of standard memory macros by integrating logic within the array, which exacerbates susceptibility to \glspl{seu} and aging effects that degrade circuit performance over time \cite{chenSpecialSessionOvercoming2024}. Consequently, standard digital accelerators remain the primary choice for mission-critical space systems, offering a baseline level of manufacturing maturity and intrinsic robustness that current in-memory alternatives cannot yet guarantee.
\section{Architecture}
\subsection{Heterogeneous RISC-V Cluster}
\label{sec:riscv_cluster}

\begin{figure}[t]
    \centering
    \includegraphics[width=0.8\columnwidth]{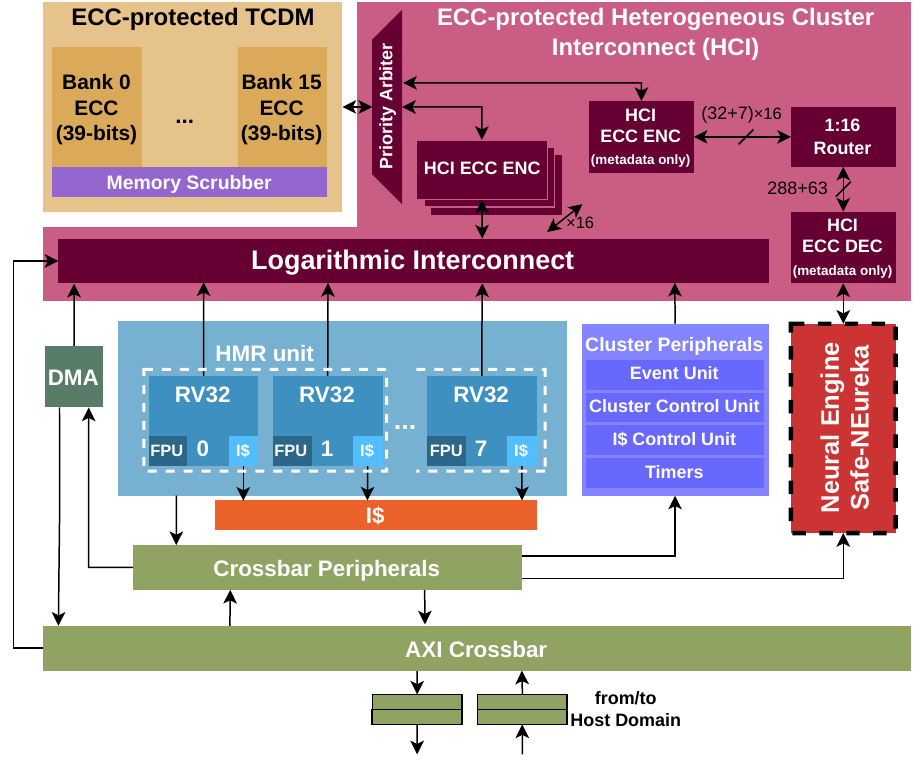}
    \caption{Multi-core RISC-V (RV32) cluster architecture augmented with the Safe-NEureka neural engine, featuring Safe-protected RISC-V cores and ECC-protected TCDM and HCI.}
    \label{cluster_archi}
\end{figure}

Our architectural baseline builds upon the open-source \gls{pulp} cluster \cite{nadalini3TOPSRISCV2023}, augmented with the general-purpose \gls{hmr} extensions proposed by Rogenmoser et al. \cite{rogenmoserHybridModularRedundancy2023}. We extend this system with domain-specific fault-tolerant acceleration, aiming to unify general-purpose reliability with the efficiency of heterogeneous computing. As depicted in Fig.~\ref{cluster_archi}, the cluster features eight RISC-V \gls{dsp} cores built on the Flex-V\footnote{https://github.com/pulp-platform/flex-v} architecture. These cores support the \texttt{RV32IMCF} instruction set with \texttt{XpulpNN-mixed} extensions and feature dedicated \glspl{fpu} alongside a hierarchical instruction cache. The cores are paired within \gls{hmr} units to enable dynamic configuration, allowing grouped execution to prioritize redundancy or ungrouped execution to maximize performance. To accelerate \gls{ai} workloads, the system integrates the Safe-NEureka accelerator (detailed in Sec.~\ref{sec:hmr_neureka}), while a \gls{dma} engine ensures efficient data transfer across the memory hierarchy.

To enhance cluster-wide reliability, the baseline \gls{tcdm} is augmented with programmable memory scrubbers for periodic error correction and a Hsiao \gls{secded} \gls{ecc} scheme~\cite{hsiaoClassOptimalMinimum1970}. Consequently, the shared 128 kB L1 \gls{tcdm} is organized into 16 banks utilizing 39-bit words, comprising 32 bits of data and 7 bits of \gls{ecc}.
Access to the \gls{tcdm} is managed by a low-latency, high-bandwidth \gls{hci} that enables seamless resource sharing. The interconnect topology differentiates by initiator: RISC-V cores and the \gls{dma} share a logarithmic interconnect, while the Safe-NEureka accelerator utilizes a dedicated, high-priority interconnect branch. The \gls{hci} extends the \gls{secded} \gls{ecc} redundancy scheme to protect all communication paths between the memory banks and processing engines. Initiator agents interact with the \gls{tcdm} through this reliable interconnect using Hsiao encoder (ENC) and decoder (DEC) pairs.

For the RISC-V cores, ECC encoding is performed via dedicated ENCs positioned after the logarithmic interconnect. Conversely, Safe-NEureka performs ECC encoding and decoding internally. Regardless of the initiator, the ENC generates check bits for both the payload and the transaction metadata (address, byte enable, and write enable). To handle the accelerator's wide data width, an additional ECC decoder operates exclusively on metadata at the accelerator interface; this allows a single 32-bit address and 288-bit-wide data channel to be split and routed into 16 memory-bank-compatible requests. Downstream ENCs subsequently reapply metadata protection before arbitration. Metadata decoding and correction occur immediately before transactions reach the \gls{tcdm} banks. During write operations, the full 39-bit payload is stored in the memory banks. During read operations, the payload traverses the fault-tolerant interconnect and is decoded only upon arrival at the destination—either the logarithmic interconnect interface or the accelerator.

\subsection{Safe-NEureka Architecture}
\label{sec:hmr_neureka}

\subsubsection{Accelerator microarchitecture}
Safe-NEureka is a parametric and run-time configurable accelerator for RISC-V multi-core clusters, designed as an extension of the open-source NEureka architecture~\cite{prasadSpecializationMeetsFlexibility2023} and following the \gls{hwpe} paradigm\footnote{https://hwpe-doc.readthedocs.io}.
Preserving the capabilities of the baseline, the accelerator supports 3$\times$3 depthwise (DW) convolutions as well as standard dense convolutions, including both 3$\times$3 spatial and 1$\times$1 pointwise (PW) types.
It operates on 8-bit activations and supports variable weight precision ranging from 2 to 8 bits.

The microarchitecture consists of three primary blocks: the \textsc{Streamer}, the \textsc{Controller}, and the \textsc{Engine}. The \textsc{Streamer} acts as a load-store unit, connected to a shared L1 memory via a high-bandwidth port offering a 288-bit payload and 63-bit \gls{ecc} per cycle. It utilizes a single \gls{tcdm} bus shared between source and sink engines, which execute load and store requests respectively.
The \textsc{Controller} contains the internal register file, a memory-mapped control port, a control \gls{fsm}, and a lightweight programmable tiling unit ($\mu$loop) designed to compute memory access offsets for \gls{dnn} nested loops.
Finally, the \textsc{Engine} serves as the computing datapath. It is subdivided into an input buffer, an input dispatching matrix, and a parametric H$\times$W array of \glspl{pe}. Each \gls{pe} includes MAC blocks alongside normalization and quantization logic and is responsible for computing the contributions to a single output pixel across 32 channels. In this work, we target a 4$\times$4 \gls{pe} configuration.

\begin{figure}[t]
    \centering
    \includegraphics[width=\columnwidth]{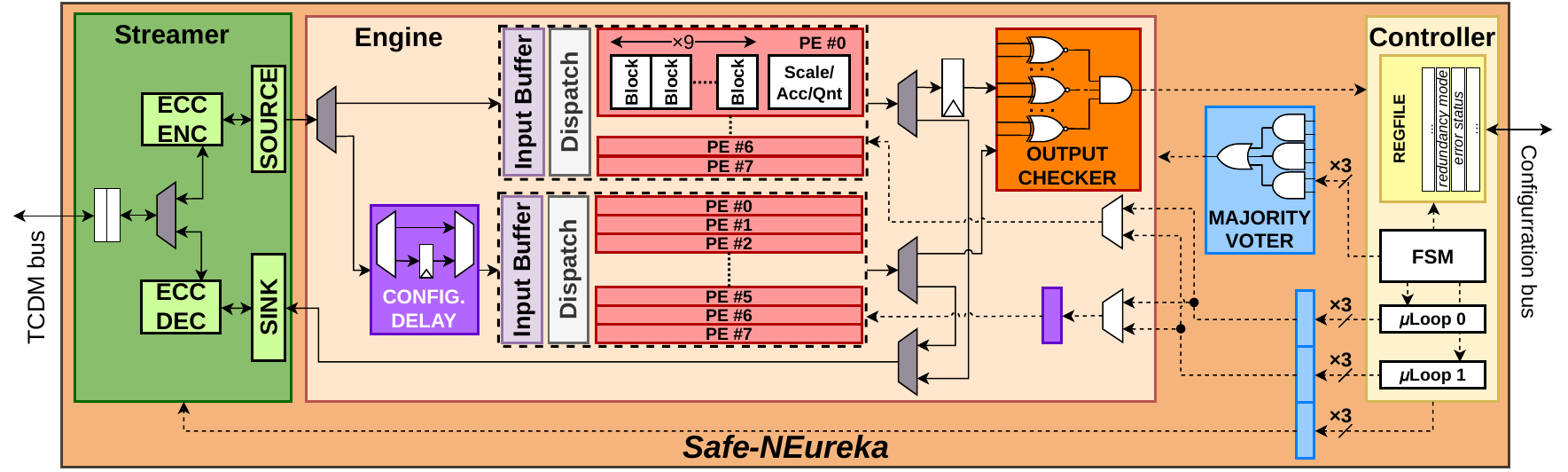}
    \caption{Architecture of run-time reconfigurable Safe-NEureka accelerator.}
    \label{HMR-NEureka_archi}
\end{figure}

\subsubsection{Hybrid Modular Redundancy}
To implement the Safe-NEureka architecture, the baseline 4$\times$4 \gls{pe} array is partitioned into two independent 4$\times$2 subarrays, as depicted in Fig.~\ref{HMR-NEureka_archi}. This partitioned datapath allows a dual-mode operation: it can function in \gls{dmr} execution, where one array redundantly executes the tasks of the other for error checking, or it can use its full throughput, leveraging both 4$\times$2 units to process separate tiles in parallel. A set of multiplexers routes inputs and outputs to direct data flow to the two halves according to the active mode.
To mitigate common-mode faults during \textit{redundancy} mode, a one-cycle temporal offset is introduced by buffering the shadow datapath inputs and the main datapath outputs before comparison. These buffers are bypassed when operating in \textit{performance} mode.
Additionally, recognizing the \textsc{Controller}'s critical role despite its small area footprint, it is protected using \gls{tmr} via triplication and majority voting, as its control signals are distributed across all \glspl{pe}.

Like the original NEureka, Safe-NEureka divides its operation into three internal stages: \textsc{input load} (filling the internal buffer), \textsc{mm} (matrix multiplication with weights streamed and broadcast to \glspl{pe}), and \textsc{streamout} of outputs.
The proposed \gls{hmr} mechanism intervenes specifically during \textsc{input load} to manage data duplication and \textsc{streamout} to perform error detection. Notably, the streaming of weights remains unaffected by the HMR mode, as weights are always broadcast identically to all PEs. This minimizes the complexity overhead, as the core \textsc{mm} compute phase requires no modification.

\begin{figure}[t]
    \centering
    \includegraphics[width=1\columnwidth]{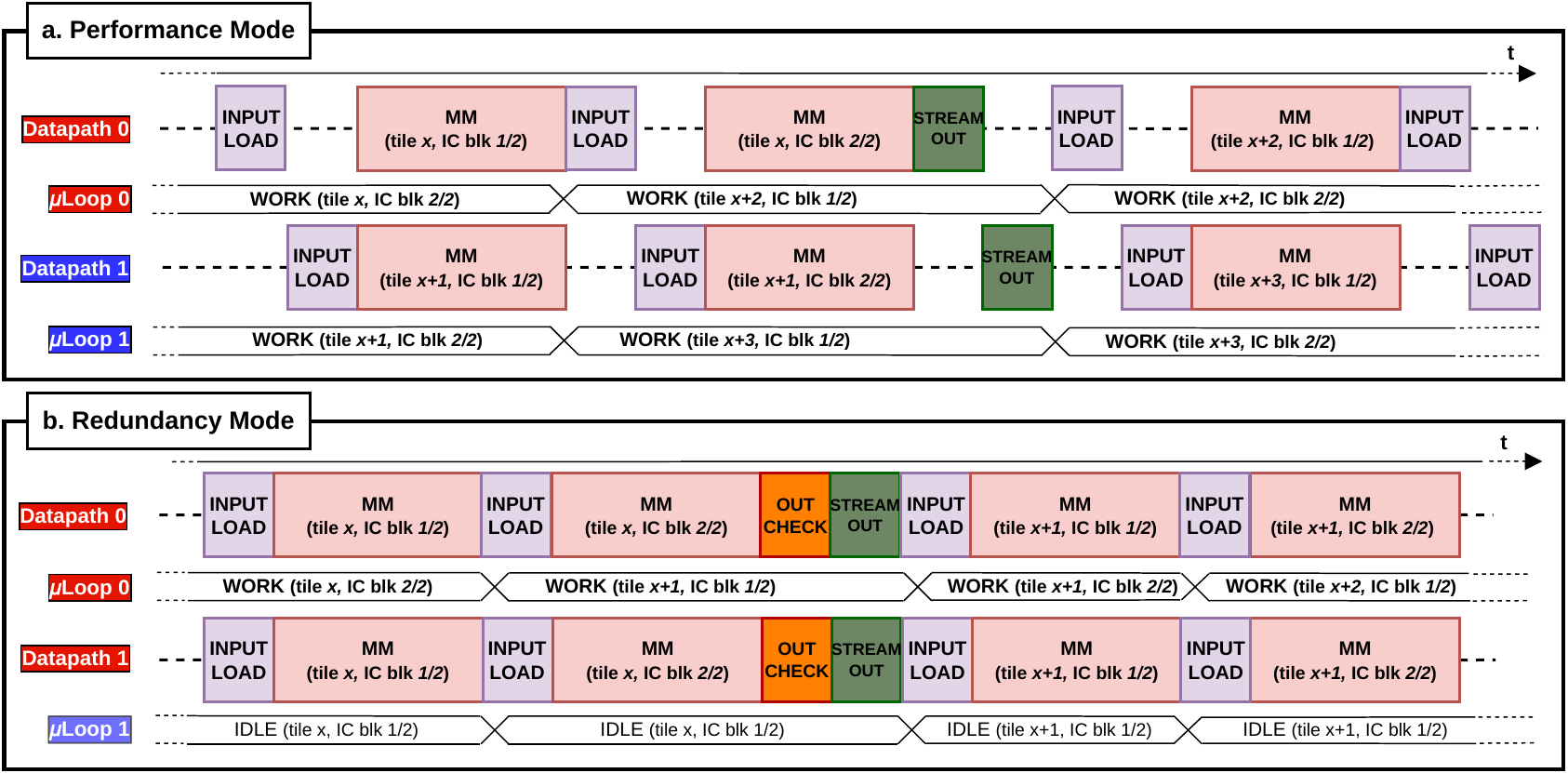}
    \caption{Comparison of datapath and $\mu$loop execution flows during fault-free operation in Safe-NEureka's \textit{performance} and \textit{redundancy} modes.}
    \label{fig:FSM-normal}
\end{figure}

Fig.~\ref{fig:FSM-normal} depicts the pipeline diagram of the controller \gls{fsm} and $\mu$loops, illustrating the state evolution and the alternation of different tiles during fault-free execution. The primary function of the $\mu$loop is to calculate memory access offsets for the four nested loops required by convolutional layers, achieved via a custom microcode. To ensure continuous data availability, the $\mu$loops calculate addresses for the Input Channel block (\texttt{IC blk}) immediately following the one currently being processed in the datapath; this ensures memory access pattern computation is finalized prior to the subsequent \textsc{Input Load} phase. The depicted scenario illustrates a kernel (e.g., pointwise or dense) with 64 input channels, which requires two iterations over the input channels—indicated as \texttt{IC blk}s—to complete the accumulation.

\begin{figure}[t]
    \centering
    \includegraphics[width=0.95\columnwidth]{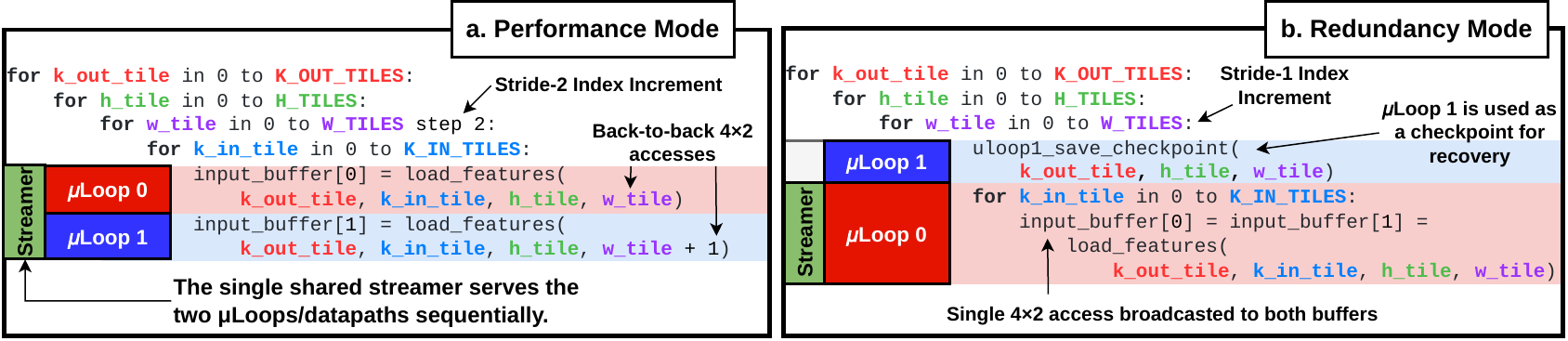}
    \caption{Pseudo-code for the Safe-NEureka $\mu$loop tiling patterns in \textit{performance} and \textit{redundancy} modes.}
    \label{fig:FSM-normal-loops}
\end{figure}

In \textit{performance} mode, the accelerator mantains a comparable throughput to a standard 4$\times$4 NEureka array. The two datapaths operate independently to process distinct input tiles simultaneously, each driven by its own $\mu$loop logical unit under a shared \textsc{controller}. As illustrated in Fig.~\ref{fig:FSM-normal-loops}a, this mode employs modified $\mu$loop microcode tailored for the two 4$\times$2 datapaths. By iterating over input row tiles with a stride of two, it effectively replicates the access patterns of the baseline 4$\times$4 array. However, unlike the baseline, a single \textsc{streamer} must service two smaller 4$\times$2 datapaths rather than a unified 4$\times$4 unit. This requires the controller to serialize the loading of input buffers and the draining of \gls{pe} accumulators for each half. Consequently, the \textsc{input load} and \textsc{streamout} FSM states incur a slight overhead due to these configuration cycles, while compute-bound states remain unaffected.

Conversely, in \textit{redundancy} mode (Fig.~\ref{fig:FSM-normal-loops}b), the $\mu$loop tiling microcode processes a single 4$\times$2 output tile. The \textsc{streamer} broadcasts identical inputs to both the main and shadow arrays, producing redundant outputs. Subsequently, an \textsc{output check} state verifies these results via bitwise XNOR-reduction before \textsc{streamout}.
To support rollback without overhead, $\mu$loop 0 acts as the active tiling unit for both datapaths, while $\mu$loop 1 serves as a passive recovery checkpoint. Specifically, $\mu$loop 1 remains pinned to the start of the current tile iteration (at the innermost input channel loop) and only advances to realign with $\mu$loop 0 upon successful verification of the outputs.

The \textsc{controller} manages runtime switching between these modes. To trigger a mode change, a cluster processor writes to a specific memory-mapped configuration field in the register file while Safe-NEureka is idle. All subsequent jobs offloaded to the accelerator execute in the newly selected mode. This process is transparent to the user application: the same configuration parameters (e.g., input/output data tile dimensions) are used regardless of the active \gls{hmr} mode.

\begin{figure}[tb]
    \centering
    \includegraphics[width=1\columnwidth]{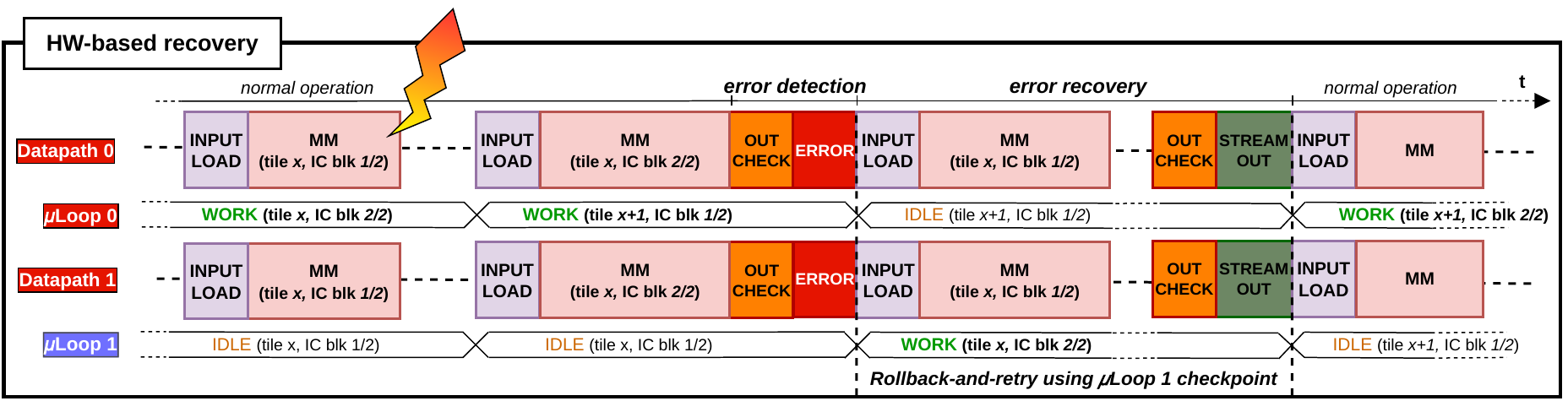}
    \caption{Graphical representation of the \textit{redundancy} mode Safe-Neureka FSM and $\mu$loops operation during error detection and subsequent recovery.}
    \label{recovery:FSM}
\end{figure}

\subsubsection{Hardware-assisted error recovery}
\label{sec:recovery}

Upon detecting a mismatch during the \textsc{streamout} phase, the checker raises an \texttt{error\_detected} flag. This signal transitions the accelerator’s \gls{fsm} into an error state, logs the event in the memory-mapped \texttt{error\_status} register, and triggers a hardware-based rollback-and-retry mechanism.

If an error is flagged, the system utilizes the preserved state of $\mu$loop 1 to execute a fast rollback, as illustrated in Fig.~\ref{recovery:FSM}. Consequently, the computation of the corrupted tile is reissued starting from the \textsc{input load} stage, effectively correcting soft errors affecting the buffers or transient faults within the datapath. This approach minimizes recovery latency to the deterministic duration required to recompute the specific tile. Furthermore, the recovery logic is robust against subsequent faults; the system maintains an iterative check where outputs are re-verified before streamout, re-triggering the rollback procedure if a mismatch persists.
These timing and reliability properties are further analyzed in Sec.~\ref{sec:results_recovery}.

\subsubsection{Error correction in \textsc{streamer}}
To compound the fault protection of the datapath, full \gls{ecc} is employed in Safe-NEureka's \textsc{streamer} in both operating modes.
Together with \gls{ecc} support in the cluster interconnect and \gls{tcdm}, this enables near-optimal protection on the entire data plane.
\gls{secded} \gls{ecc} encoders and decoders are employed.
Within the \textsc{streamer}, the high-bandwidth 288-bit data payload connected to Safe-NEureka's memory port is partitioned into nine separate 32-bit chunks, each independently encoded and decoded with seven additional check bits for a total of 351 bits. Detected errors are recorded in dedicated, memory-mapped, software-accessible registers in the register file, enabling error tracking.
\section{Experimental Results}




We implemented Safe-NEureka in synthesizable SystemVerilog HDL. RTL performance evaluations were conducted using QuestaSim, with data collected via internal performance counters and simulation traces. For physical implementation, we targeted GlobalFoundries 12 nm technology. Synthesis was performed using Synopsys Design Compiler 2020.09 at a target clock frequency of 500 MHz (SSPG corner, 0.72 V, -40 °C), followed by Place \& Route using Cadence Innovus 21.17 under the same constraints. Power analysis was carried out using Synopsys PrimePower 2022.03 in the TT corner (0.80 V, 25 °C). Finally, we assessed fault tolerance by performing fault injection on the post-synthesis netlist using the Synopsys VC-Z01X 2025.06 framework.

\subsection{Power, Performance and Area evaluation}


\begin{figure}[t]
    \centering
    \includegraphics[width=0.55\columnwidth]{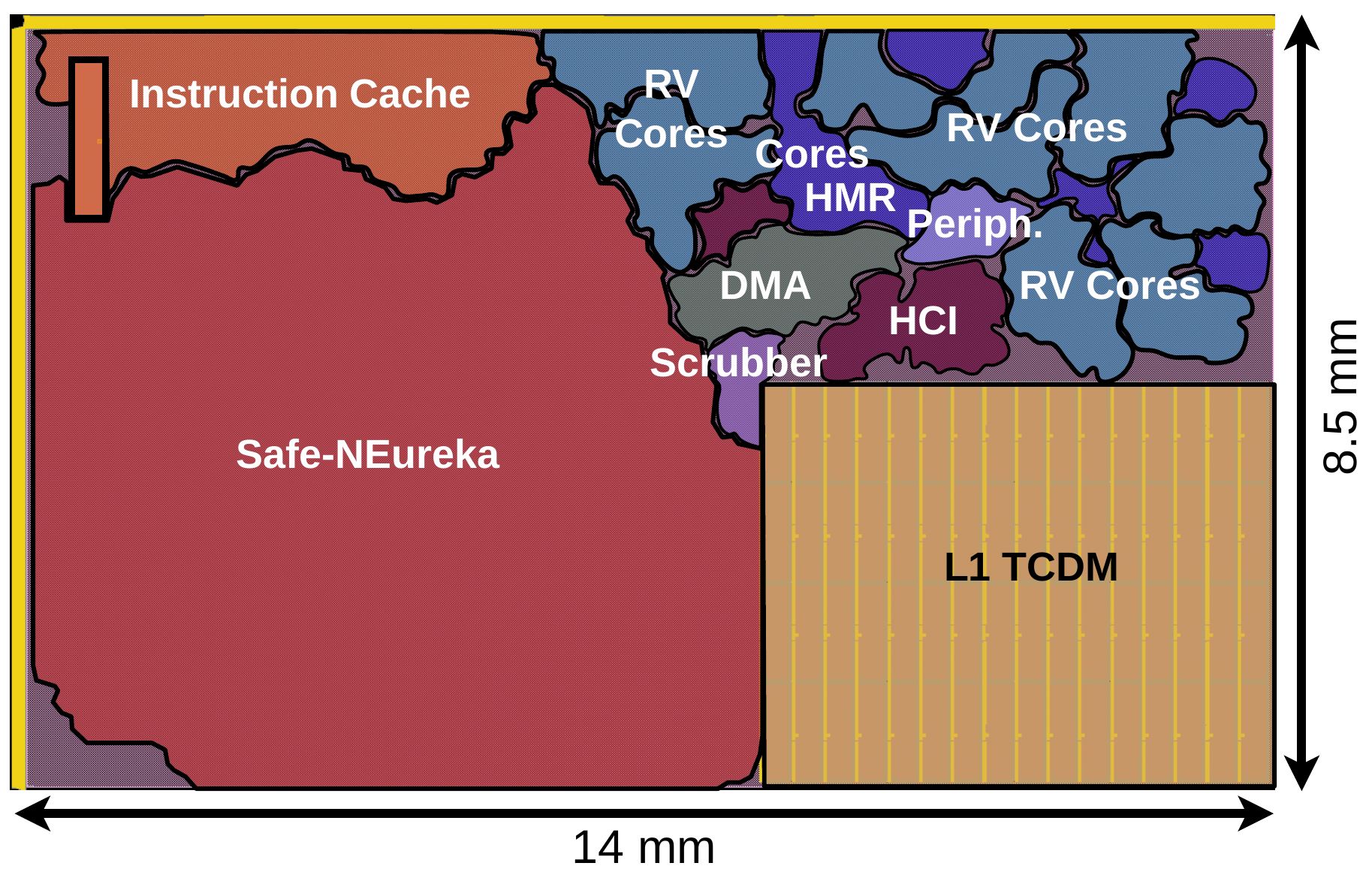}
    \caption{Placed-and-routed layout of the cluster featuring Safe-NEureka, ECC-protected TCDM, HCI, and HMR-protected RISC-V cores.}
    \label{post_layout:layout}
\end{figure}

\begin{figure}[t]
    \centering
    \includegraphics[width=\columnwidth]{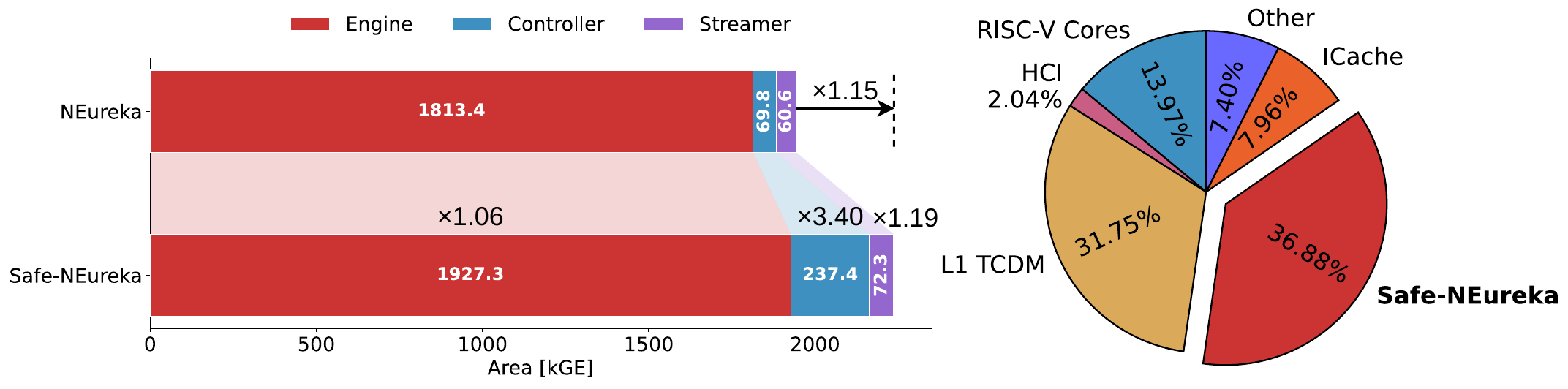}
    \caption{Area breakdown of the Safe-NEureka accelerator compared to the NEureka baseline (left), and percentage area distribution of the heterogeneous cluster (right).}
    \label{post_layout:breakdown}
\end{figure}

The full cluster with Safe-NEureka has been implemented as a macro of \SI{14}{\milli\meter}$\times$\SI{8.5}{\milli\meter}, with a density of 55\%. Fig.~\ref{post_layout:layout} shows the post-layout cluster implementation highlighting the main components. As detailed in the area breakdown in Fig.~\ref{post_layout:breakdown}, Safe-NEureka occupies a significant portion of the total cluster design ($\sim$36.9\%).

The proposed Safe-NEureka occupies a total area of approximately 0.271 mm\textsuperscript{2}. Compared to the baseline NEureka area of 0.235 mm\textsuperscript{2} (excluding \gls{hmr}, \textsc{streamer} \gls{ecc}, and \textsc{controller} \gls{tmr}), this represents an area overhead of roughly 15\% at the individual IP level and less than 6\% at the cluster level. As detailed in Fig.~\ref{post_layout:breakdown}, the \textsc{engine} area increases by 6\% due to the additional hardware required for dual operating modes, output checkers, and temporal delay buffers. Although full TMR protection and the increased control logic complexity of the duplicated $\mu$loop increase the \textsc{controller} area by 240\%, this module constitutes only $\sim$10\% of the total Safe-NEureka area. Consequently, the critical reliability gains justify the cost of protection via replication. Finally, the \textsc{streamer} area increases by 19\% with the inclusion of ECC support. Nonetheless, the datapath remains the dominant contributor to overall area utilization ($\sim$86\%).

To evaluate the performance, power, and energy overheads of both the \textit{redundancy} and \textit{performance} operating modes of Safe-NEureka relative to the baseline NEureka, we employed single convolutional layer benchmarks targeting 8-bit weight precision. All extracted figures refer to the Safe-NEureka integrated inside the heterogeneous cluster, where the pointwise ($[k_{i},k_{o},h_{o},w_{o}]=[256,64,16,16]$), dense ($[256,32,8,8]$) and depthwise ($[256,256,8,8]$) convolutions were sized specifically to fit the available 128\,kiB L1 TCDM. Here, $k$ denotes channels (input/output) and $h_{o}/w_{o}$ denote the spatial height and width of the output feature map. The region of interest for the profiling is tuned on the window where the accelerator is actively working, i.e. it does not feature its programming.

\begin{figure}[t]
    \centering
    \includegraphics[width=1\columnwidth]{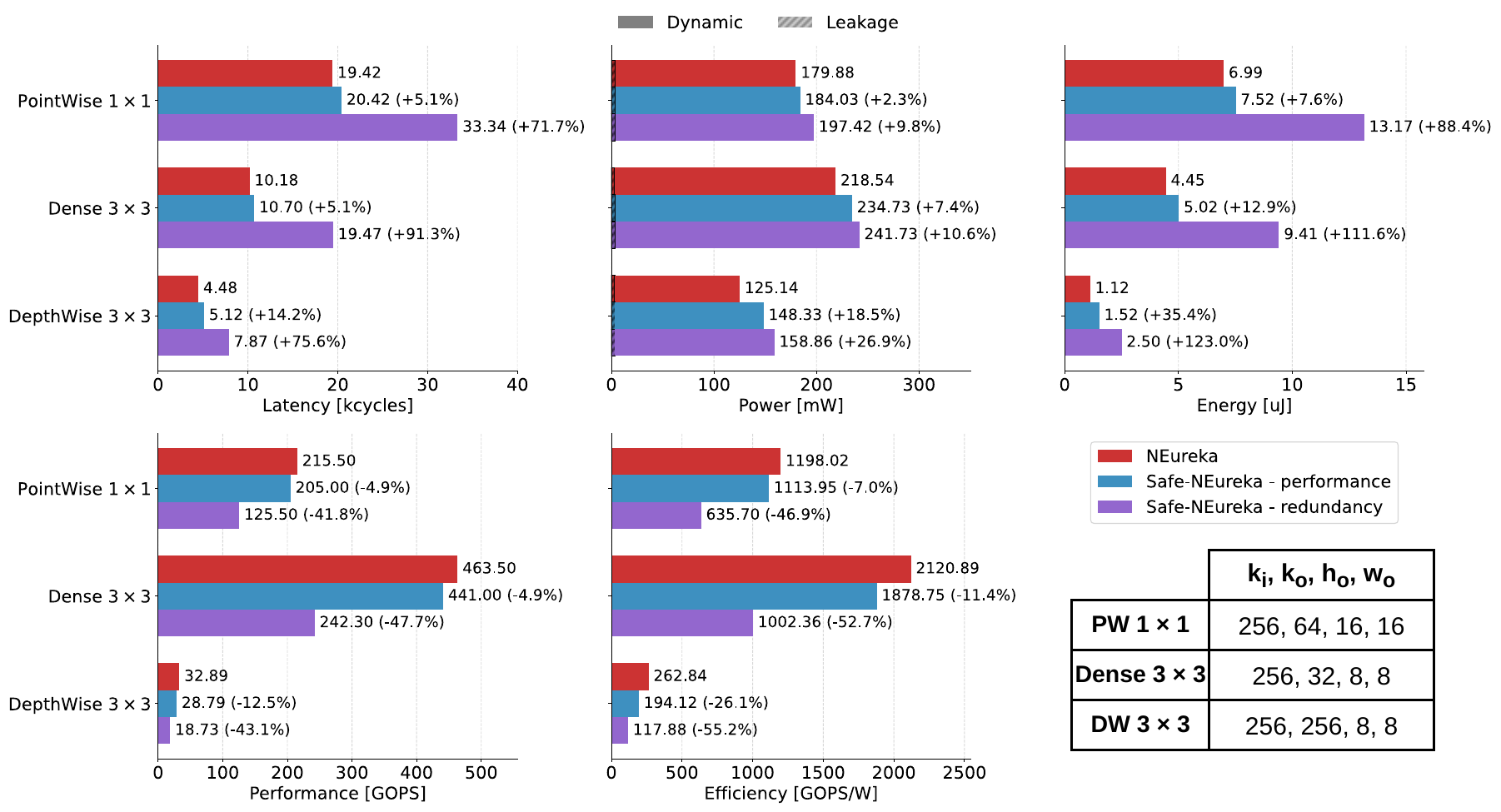}
    \caption{Latency, power, energy, and efficiency of Baseline NEureka versus Safe-NEureka (in Performance and Redundancy modes) across pointwise 1$\times$1, dense 3$\times$3, and depthwise 3$\times$3 layers.}
    \label{fig:power_perf}
\end{figure}

Fig.~\ref{fig:power_perf} summarizes the workload metrics. The results indicate that Safe-NEureka's \textit{performance} mode incurs minimal latency overhead for pointwise and dense convolutions ($\sim$5\% degradation) and remains limited for depthwise layers ($\sim$13\%) compared to the NEureka baseline. This cost is primarily attributed to the additional control mechanisms required to manage the dual datapaths. Notably, in \textit{redundancy} mode, execution time increases sub-linearly by 70\% to 90\%, depending on the operating mode.
As expected, power consumption in Safe-NEureka increases relative to the baseline due to the \gls{tmr} protection active in both operating modes. However, for the most compute-dense kernel (3$\times$3 dense convolution), this power overhead remains below 8\%. \textit{Redundancy} mode consumes moderately more power than \textit{performance} mode, largely due to the switching activity of the additional reliability hardware such as time-delay buffers and checkers. Nevertheless, also in \textit{redundancy} mode there is full utilization of \glspl{pe}, which are fed identical inputs.
Energy consumption, which accounts for both power and latency overheads, increases by 88\% to 123\% in \textit{redundancy} mode depending on the benchmark due to the combined effect of longer execution times and higher power draw. Regarding efficiency, the NEureka baseline peaks at 2.1 TOPS/W for dense 3$\times$3 convolution. Safe-NEureka sees a limited reduction to 1.9 TOPS/W (-11.4\%) in \textit{performance} mode, decreasing further to 1.0 TOPS/W (-52.7\%) in \textit{redundancy} mode. This differential highlights the inherent trade-off between prioritizing maximum throughput versus fault tolerance.

\begin{figure}[t]
    \centering
    \includegraphics[width=1\columnwidth]{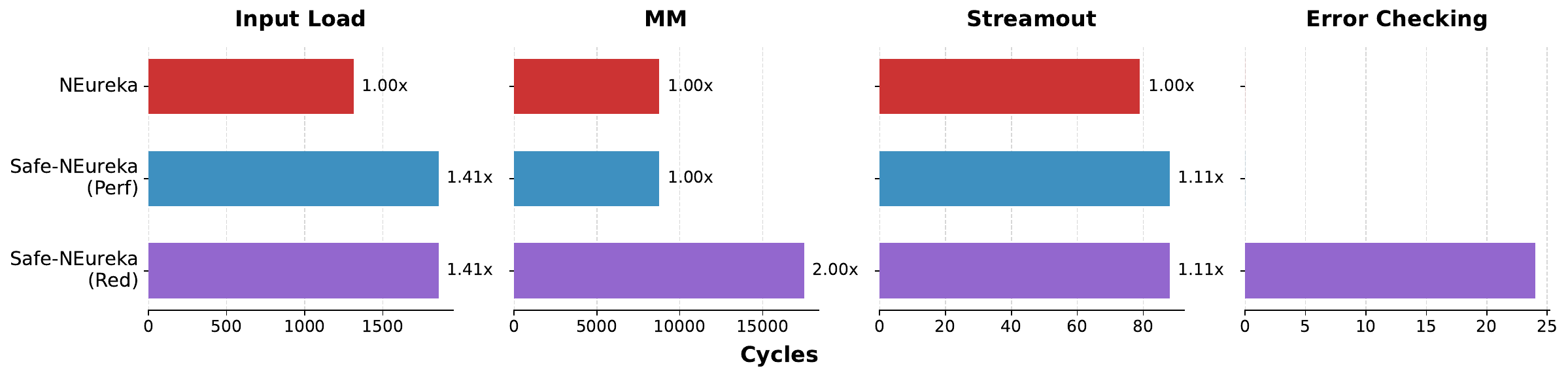}
    \caption{Cycle breakdown by controller FSM states for NEureka and Safe-NEureka (operating in performance and redundancy modes). States are categorized by macro operation: \textsc{Input Load}, \textsc{MM}, \textsc{Streamout}, and Error Checking. The evaluated layer configuration is $[k_{i},k_{o},h_{o},w_{o}]=[256,32,8,8]$ performing 3$\times$3 dense convolution.}
    \label{fig:fsm_breakdown}
\end{figure}

To explain the latency overheads and their sublinear latency scaling, Fig. \ref{fig:fsm_breakdown} details the cycle distribution of the Safe-NEureka controller \gls{fsm} across its macro operations—\textsc{Input Load}, \textsc{MM}, \textsc{Streamout}, and redundancy checks—for a dense convolution on layer $[k_{i},k_{o},h_{o},w_{o}]=[256,32,8,8]$. Notably, the total \textsc{Input Load} cycles remain constant across both operating modes, though for distinct architectural reasons. In \textit{performance} mode, loading a tile requires two back-to-back operations to service the split 4$\times$2 datapaths. Conversely, in \textit{redundancy} mode, the accelerator operates as a single 4$\times$2 datapath (shadowed by a redundant copy), necessitating double the load iterations to process the same data volume. This doubling factor, compounded by the reconfiguration latency required to program the streamer, results in a 41\% overhead compared to the baseline NEureka. Likewise, the \textsc{streamout} operations incur an 11\% penalty due to similar configuration latencies.
\textsc{mm} cycles remain unchanged in \textit{performance} mode relative to the baseline, as the \glspl{pe} array of NEureka and Safe-NEureka operate identically. Conversely, \textit{redundancy} mode exhibits an exact doubling of cycles, consistent with the expected halving of effective parallelism. Finally, the error checking cycles represent a minimal fraction of the overall execution, quantified as $(2 + \text{timeshift cycles}) \times \text{number of streamouts}$, amounting to just 24 cycles for the considered example.

\subsection{Full-network performance evaluation}
\label{sec:perf}

\begin{figure}[t]
    \centering
    \includegraphics[width=0.9\columnwidth]{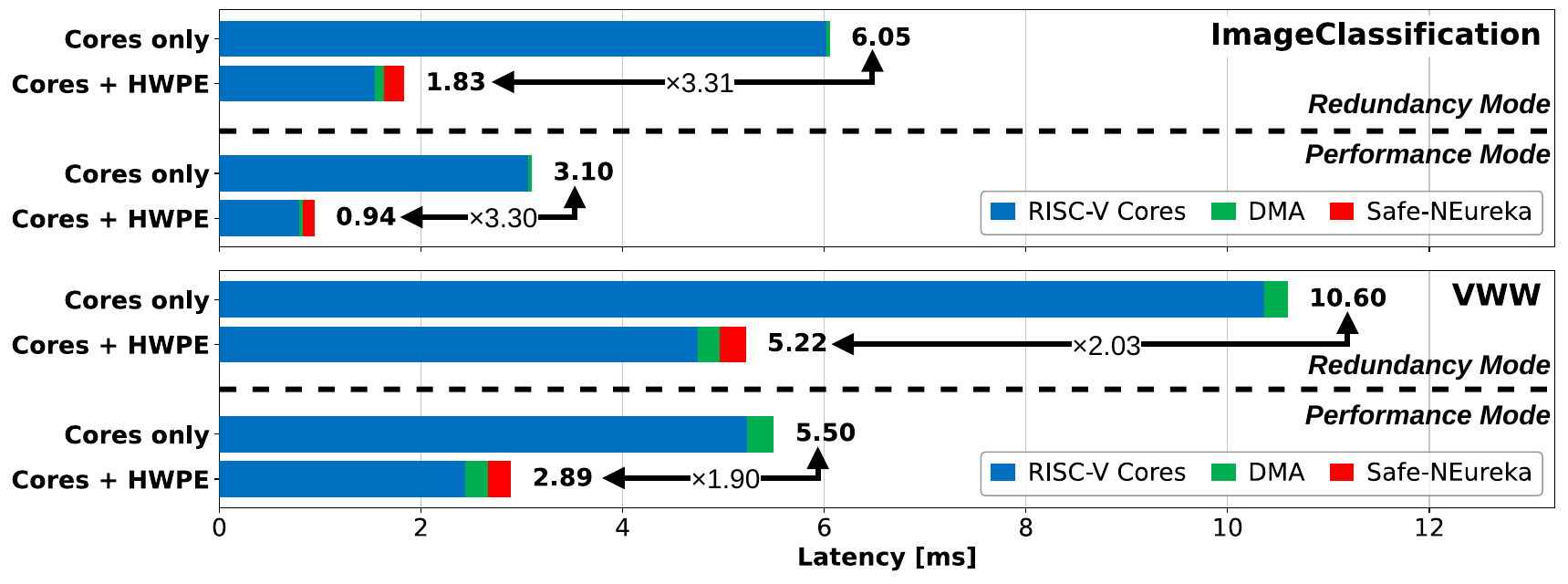}
    \caption{End-to-end execution latency at \SI{500}{\mega\hertz} for the MLPerf Tiny \textit{VisualWakeWords (VWW)} and \textit{ImageClassification} benchmarks. The figure compares the baseline 8-core HMR RISC-V cluster against the Safe-NEureka enhanced clsuter operating in both \textit{performance} and \textit{redundancy} modes.}
    \label{fig:full_network}
\end{figure}

To extend benchmarking to full networks, we utilized the event-driven GVSoC simulator~\cite{bruschiGVSoCHighlyConfigurable2021}, configured with a parametric model of NEureka.
GVSoC provides timing accuracy within $\sim$10\% of hardware while maintaining simulation speeds orders of magnitude faster than cycle-accurate alternatives.
We simulated the cluster from Sec.~\ref{sec:riscv_cluster} using two MLPerf Tiny networks: \textit{VisualWakeWords} (MobileNet-V1 with input image size 96$\times$96) and \textit{ImageClassification} (ResNet-8).
We employed the Deeploy framework~\cite{schererDeeployEnablingEnergyEfficient2024} to convert these ONNX models into ANSI C code, generating complete tiling schedules with double-buffered DMA transfers.

Fig.~\ref{fig:full_network} presents the execution latency at \SI{500}{\mega\hertz}, including control overheads and non-overlapped DMA transfers.
We compared the execution on 8 RISC-V cores featuring HMR~\cite{rogenmoserHybridModularRedundancy2023} (in either \textit{performance} or \textit{redundancy} mode) against the same cluster setup enhanced with Safe-NEureka acceleration operating in the same mode of the cores.
\textit{VisualWakeWords} executes in \SI{2.89}{\milli\second} (\SI{5.22}{\milli\second}) in \textit{performance} (\textit{redundancy}) mode, while \textit{ImageClassification} completes in \SI{0.94}{\milli\second} (\SI{1.83}{\milli\second}).
In terms of relative improvement, Safe-NEureka accelerates the pointwise 1$\times$1 layers in \textit{VisualWakeWords}, yielding end-to-end speedups of 1.90$\times$ and 2.03$\times$ in \textit{performance} and \textit{redundancy} modes, respectively, compared to their cores-only counterparts.

\subsection{Fault-coverage}

To evaluate the fault-tolerance capabilities of the proposed architecture, we conducted a hardware-based fault-injection simulation. Hardware-based simulations, particularly at the gate level, offer a balance between accuracy and feasibility for evaluating resilience~\cite{ruospoSurveyDeepLearning2023, ahmadilivaniSystematicLiteratureReview2024}. In this work, we focus on injecting \gls{set} and \gls{seu}-like faults into the post-synthesis netlist of Safe-NEureka. \glspl{seu} involve flipping one bit within storage elements, while \glspl{set} are simulated by flipping a victim combinational node for one cycle between two consecutive negative clock edges to ensure the presence of a sampling edge.

The experimental setup considers both a 1$\times1$ pointwise and a 3$\times$3 convolutional kernel, with a $[k_{i},k_{o},h_{o},w_{o}]=[64,64,8,8]$ layer configuration for both. Since gate-level fault injection tests are inherently time-consuming, smaller layers were chosen relative to those in Sec.~\ref{sec:perf} to balance representative sizing with computational constraints. From the full set of pairs defined by injection times and target locations (encompassing all sequential elements and combinational nets), a random subset of 100,000 faults was selected. Following a Monte Carlo approach, each simulation injects a single fault to monitor the system's response, such as recovery activation or memory corruption, and classifies the final outcome.

\begin{figure}[t]
    \centering
    \includegraphics[width=0.8\columnwidth]{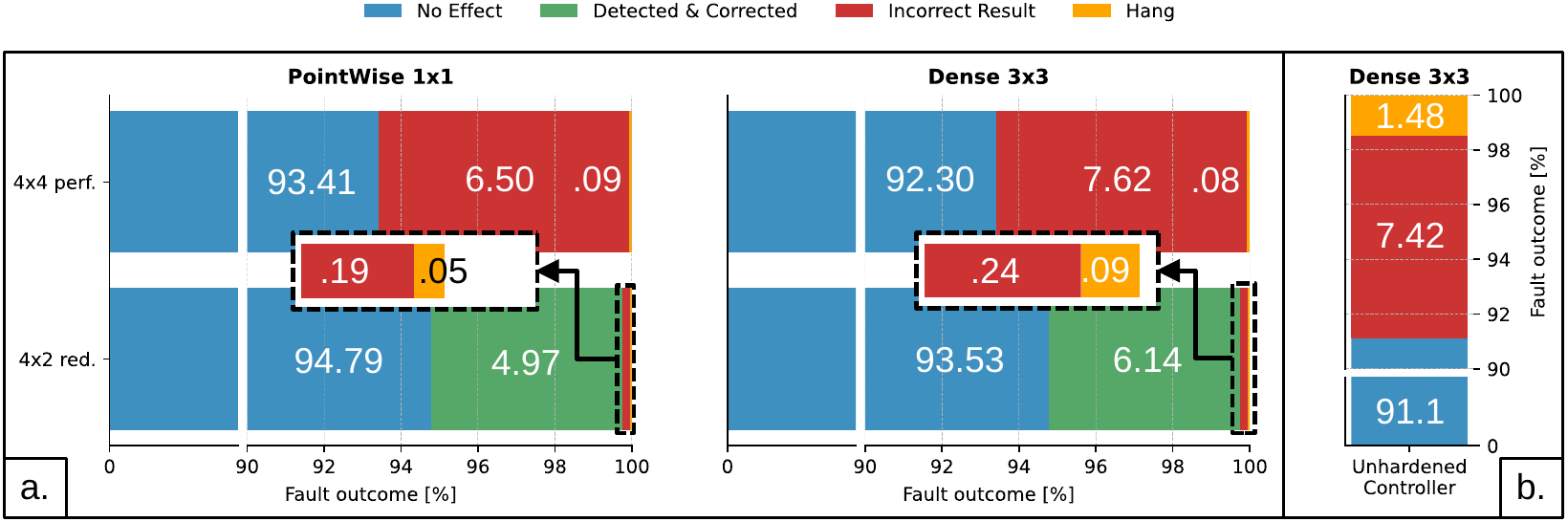}
    \caption{Distribution of fault injection outcomes (100,000 injections per scenario). 
(a) NEureka and Safe-NEureka on 1$\times$1 pointwise and 3$\times$3 dense convolution layers. 
(b) NEureka on 3$\times$3 dense convolution with faults injected exclusively into the unprotected controller. 
The outcome axis includes a break to highlight the 90\%--100\% interval due to the high prevalence of ineffective or masked faults.}
    \label{fig:fi}
\end{figure}

We categorized the outcomes into four types: \textit{Incorrect Result} if the computation ended with a wrong result propagated to memory; \textit{Hang} if the accelerator reached an unrecoverable internal state requiring a soft reset; \textit{Detected \& Corrected} if the fault was detected in \textit{redundancy} mode and successfully corrected; and \textit{No Effect} if the injected fault produced no noticeable effect on execution due to logical masking.
Fig.~\ref{fig:fi}a compares the fault injection results of the unhardened NEureka baseline with Safe-NEureka operating in \textit{redundancy} mode. Although the baseline exhibits a high percentage of \textit{No Effect} faults, it still reports an \textit{Incorrect Result} rate between 6.50\% and 7.60\%. The higher memory corruption rate observed in 3$\times$3 convolutions is attributed to increased datapath utilization during dense convolution workloads; consequently, the probability that a random fault corrupts active logic increases. In contrast, Safe-NEureka's \textit{redundancy} mode lowers this failure rate to approximately 0.20\%--0.25\%, representing a substantial improvement in fault tolerance by eliminating $\sim$96\% of problematic cases. Furthermore, tests demonstrate that the recovery procedure is highly effective; whenever an error is detected, it is successfully corrected.

Notably, in the baseline NEureka, observable faults within the \textsc{controller} predominantly result in \textit{Incorrect Results} rather than \textit{Hangs}. This observation is detailed in Fig.~\ref{fig:fi}b, which reports the outcomes of 100,000 \glspl{seu} injected into the unprotected \textsc{controller} (i.e., lacking \gls{tmr}) while executing a dense 3$\times$3 convolution. This behavior occurs because faults primarily cause the system to skip operations or processing phases rather than deadlock—for example, by corrupting a counter and prematurely reaching the expected end of an iteration.
Additionally, the extensive use of synchronous clears generated by the main accelerator \textsc{Controller}—a mechanism also implemented in the baseline NEureka to ensure deterministic datapath states between processing phases—provides the secondary benefit of reducing latent faults in both versions.

Of the remaining uncorrected failures, a minor portion (0.5-0.9\%) consists of system hangs. These could be mitigated through complementary cluster-level mechanisms, such as a watchdog timer programmed by the core during task offloading. The remaining failures that propagate to memory stem from faults directly affecting the \gls{tmr} voters or \gls{ecc} logic. However, these components represent a limited vulnerability surface that can be further secured through circuit-level hardening~\cite{banSimpleFaulttolerantDigital2010}.

\subsection{Recovery Performance}
\label{sec:results_recovery}

\begin{figure}[tb]
    \centering
    \includegraphics[width=0.7\columnwidth]{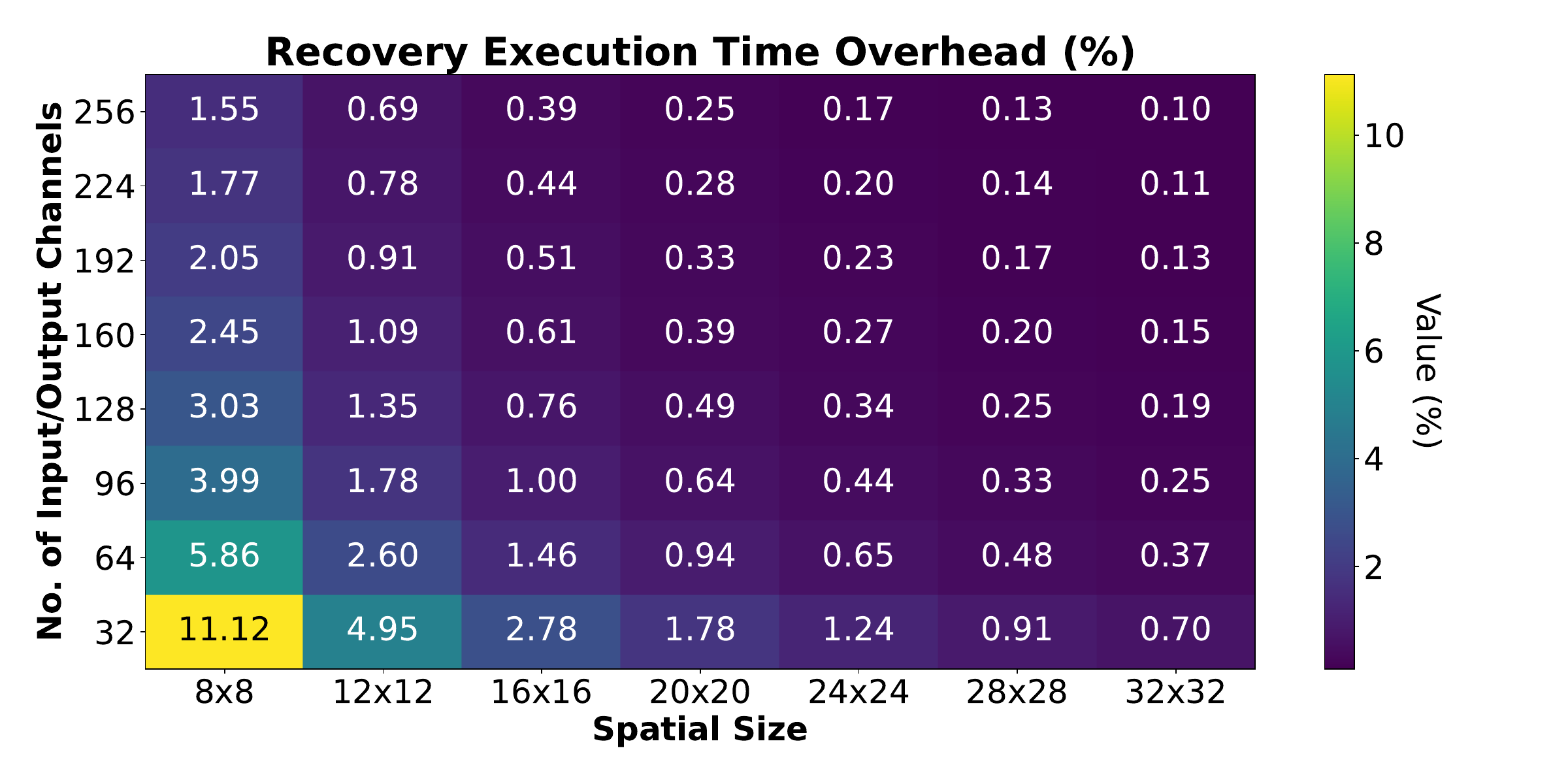}
    \caption{Percentage execution time overhead of a single detected and corrected error in a 1$\times$1 Pointwise kernel with increasing input kernel size.}
    \label{overhead}
\end{figure}

The hardware-based recovery routine described in Sec.~\ref{sec:recovery} requires a fixed number of cycles to execute in \textit{redundancy} mode, depending solely on the input channel dimension. The recovery time follows a piecewise stepwise pattern, increasing by 60 cycles for every 32 additional input channels, starting from an initial 90-cycle duration for channel counts between 1 and 32. Notably, the recovery time is independent of the number of input feature map rows and columns. As a result, the execution time overhead with respect to the fault-free case falls as the input $h_{i}$ and $w_{i}$ dimensions increase, as represented in Fig.~\ref{overhead}.

%
%

For instance, in the fault injection tests conducted on a pointwise workload with an input size of 8$\times$8$\times$64, the execution time overhead due to a single fault is +5.86\%. However, when increasing the kernel input size to 28$\times$28$\times$64, this overhead drops significantly to just +0.48\%. This trend indicates that larger spatial dimensions reduce the relative impact of recovery time on overall execution. Similar considerations apply to the dense 3$\times$3 kernel. In this case, the recovery cycle count increases by 300 units for every additional 32 input channels, starting from a baseline of 330 cycles for 32-input-channel tiles. Meanwhile, the total percentage overhead follows an inverse dependency with input spatial size, similar to that observed in pointwise operations.

\subsection{Comparison with software-based error recovery}

\begin{figure}[tb]
    \centering
    \includegraphics[width=0.65\columnwidth]{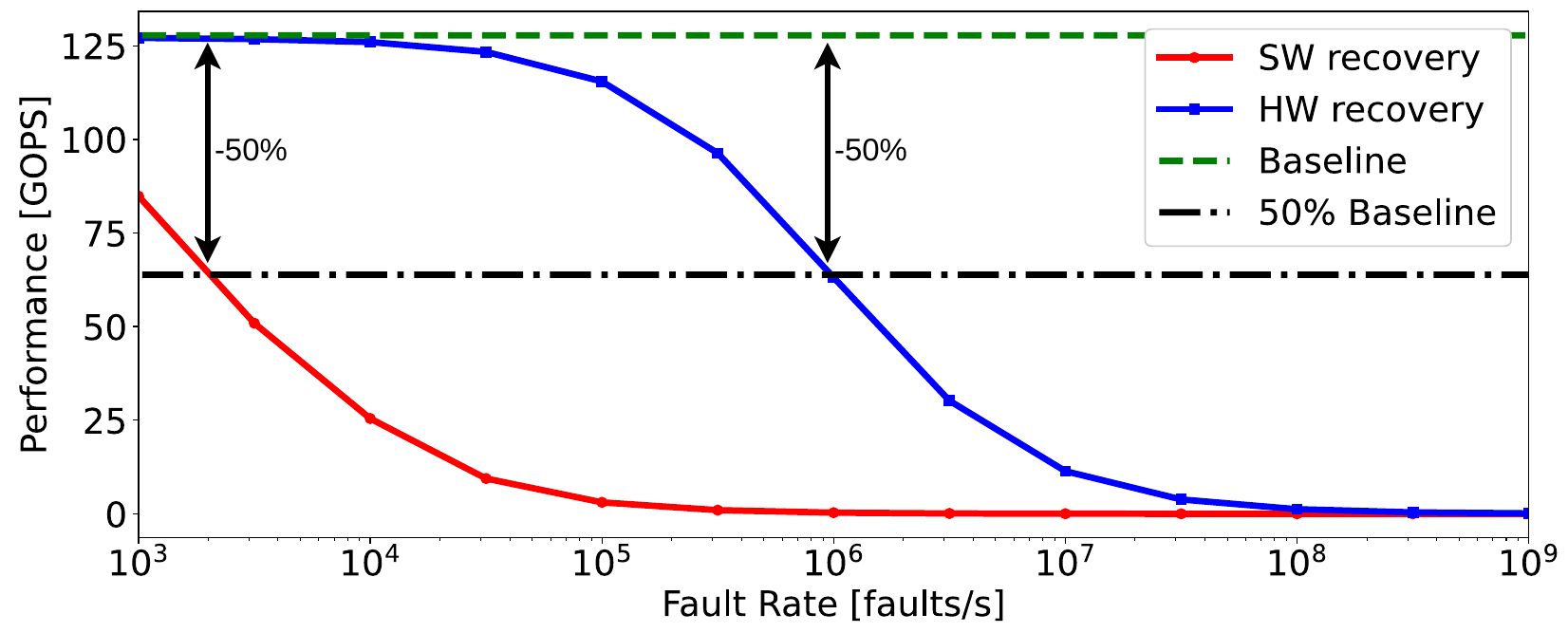}
    \caption{\textit{Redundancy} mode Safe-NEureka performance degradation at increasing fault rate; comparison between SW-based and HW-based recovery.}
    \label{multi_fault}
\end{figure}
To assess the effectiveness of the proposed hardware-assisted error recovery mechanism, we compare it with a software-based approach.
In the latter case, a detected error triggers a recovery by clearing the accelerator's state and reissuing the entire job from the start: as a consequence, recovery performance is highly unpredictable, as the overhead depends on the level of job completion at the time of the fault.
Moreover, in the case of multiple consecutive faults, performance degradation accelerates rapidly, as illustrated in Fig.~\ref{multi_fault}, which shows the impact of increasing fault rates during the execution of a benchmark application (the 1$\times$1 kernel on $[k_{i},k_{o},h_{o},w_{o}]=[256,256,28,28]$ layer). 
Since software recovery requires discarding all work completed prior to the fault, the overhead depends heavily on fault timing. Therefore, our analysis assumes an average scenario where the fault occurs midway through the job.
The proposed hardware-based rollback-and-retry recovery scheme maintains near-optimal performance with minimal degradation up to approximately $1\times10^5$ faults/s, while the software-based one starts struggling after a few consecutive injected faults. Notably, with the hardware-based scheme, a 50\% performance drop occurs only at ~$1\times10^6$ faults per second, whereas the same level of degradation is observed at just ~$2\times10^3$ faults per second with the software-based approach.
\section{Comparison with State-of-the-Art}

We quantitatively compare Safe-NEureka with \gls{soa} fault tolerance methodologies for \gls{dnn} accelerators in Tab. \ref{tab:soa}. While a comparison with ASIC-based designs is most relevant to the scope of this work, the majority of open literature targets FPGAs due to their \gls{cots} availability and \gls{hls} support. We include the open-source design of \citet{prasadSpecializationMeetsFlexibility2023} to provide a reference of a performance-optimized, unhardened accelerator, with the reported figures obtained by integrating the open source design into our experimental setup.

Methodologies such as FSA \cite{zhaoFSAEfficientFaulttolerant2022} and HyCA \cite{liuHyCAHybridComputing2022} mitigate permanent stuck-at faults by utilizing redundant computing units (\glspl{ru} or \glspl{dppu}) to dynamically bypass faulty \glspl{pe}. However, a critical limitation persists: both rely on offline or periodic identification (e.g., via \gls{bist}), rendering them too slow for the real-time detection required for runtime \gls{see} mitigation. Additionally, system performance degrades once faults exceed the available redundant capacity. In terms of overhead, FSA simulations (using SCALE-Sim and CACTI) project up to a 20\% latency penalty, though power-gating faulty \glspl{pe} yields a 20--30\% net energy reduction. The reported area increase is 3.1\%, however an actual implementation beyond SCALE-Sim is not provided. Conversely, HyCA is validated via a TSMC 40nm ASIC implementation with a footprint of 7.5 mm\textsuperscript{2} ($<$2\% overhead), though it lacks explicit quantification of power and energy consumption.

In contrast, HAp-FT \cite{weiHApFTHybridApproximate2024} mitigates both \glspl{see} and permanent faults by implementing approximate consistency checks. However, this threshold-based detection specifically exploits the inherent resilience of floating-point mantissas, where bit-flips yield significantly smaller deviations than those in the exponent. Consequently, this approach is less effective for the quantized integer formats prevalent in edge \gls{ai}, where bit-flips typically cause higher value deviations and result in lower reported recovery rates. Furthermore, similarly to FSA, HAp-FT rely on high-level models (e.g., SCALE-Sim, CACTI) rather than hardware synthesis, leaving their reported area overheads theoretical.

Masking techniques attempt to minimize hardware overhead by exploiting data characteristics or idle time. \citet{salamiResilienceRTLNN2018} propose a hybrid masking technique relying on statistical heuristics (e.g., assuming correlation between Sign and MSB bits) rather than deterministic correction. This makes reliability strictly dependent on data distribution and quantization format. Furthermore, the authors do not quantify the specific area and power overheads of their mitigation logic, relying instead on estimates from related literature.

Similarly, the opportunistic detection strategy employed by \citet{liSoftErrorMitigation2020} is strictly coupled to the utilization patterns of the model. Unlike a deterministic approach, their reliance on idle cycles implies that dense workloads would necessitate throughput throttling to maintain detection coverage. Furthermore, their mechanism relies on error masking by forcing faulty \glspl{pe} outputs to zero. While this prevents large-magnitude errors, it fails to recover the correct result, leading to inevitable accuracy degradation. Finally, while their FPGA evaluation indicates low hardware overheads ($<$17\% LUTs, $<$9\% FFs), the study lacks power and energy characterization, preventing the quantification of the actual dynamic cost of activating resources during otherwise idle periods.

\begin{table}[t]
\centering
\caption{Comparison between the state of the art and this work.}
\label{tab:soa}
\resizebox{\textwidth}{!}{%
\begin{tabular}{llllllllll}
\hline
\textbf{Identifier} & \textbf{\begin{tabular}[c]{@{}l@{}}Fault Tolerance\\ Methodology\end{tabular}} & \textbf{\begin{tabular}[c]{@{}l@{}}Fault Model /\\ Assessment\end{tabular}} & \textbf{\begin{tabular}[c]{@{}l@{}}Target\\ Implementation\end{tabular}} & \textbf{\begin{tabular}[c]{@{}l@{}}Area /\\ Resources\end{tabular}} & \textbf{Benchmark} & \textbf{Power} & \textbf{Energy} & \textbf{Latency} & \textbf{Efficiency} \\ \hline
\multirow{2}{*}{Aphelios \cite{sunApheliosSelectiveLockstep2025}} & \multirow{2}{*}{\begin{tabular}[c]{@{}l@{}}DMR (Control, Data Load),\\ partially redundant systolic array\end{tabular}} & \multirow{2}{*}{\begin{tabular}[c]{@{}l@{}}SEU /\\ Cycle-acc. FI\end{tabular}} & \multirow{2}{*}{\begin{tabular}[c]{@{}l@{}}RTL / ASIC\\ Silvaco 15nm\end{tabular}} & \multirow{2}{*}{\begin{tabular}[c]{@{}l@{}}3.03 mm² /\\ +30.6\%\end{tabular}} & ResNet-18 & N.A. & +33\% & +11\% & N.A. \\ \cline{6-10} 
 &  &  &  &  & Yolov3 & N.A. & +33\% & +11\% & N.A. \\ \hline
\multirow{2}{*}{FSA \cite{zhaoFSAEfficientFaulttolerant2022}} & \multirow{2}{*}{\begin{tabular}[c]{@{}l@{}}Recomputation modules\\ replacing faulty PEs\end{tabular}} & \multirow{2}{*}{\begin{tabular}[c]{@{}l@{}}Stuck-at /\\ Weight flipping\end{tabular}} & \multirow{2}{*}{\begin{tabular}[c]{@{}l@{}}SCALE-Sim,\\ CACTI\end{tabular}} & \multirow{2}{*}{\begin{tabular}[c]{@{}l@{}}+3.1\%\\ (analytical)\end{tabular}} & ResNet-50 & -11\% to -8\% & -30\% to -20\% & \textless{}20\% & N.A. \\ \cline{6-10} 
 &  &  &  &  & AlexNet & -11\% to -8\% & -30\% to -20\% & $\sim$0\% & N.A. \\ \hline
HyCA \cite{liuHyCAHybridComputing2022} & \begin{tabular}[c]{@{}l@{}}Recomputation modules\\ replacing faulty PEs\end{tabular} & \begin{tabular}[c]{@{}l@{}}Stuck-at /\\ Random Faulty\\ PE mapping\end{tabular} & \begin{tabular}[c]{@{}l@{}}RTL / ASIC\\ TSMC 40nm\end{tabular} & \begin{tabular}[c]{@{}l@{}}7.5 mm2 /\\ \textless 3\%\end{tabular} & \begin{tabular}[c]{@{}l@{}}AlexNet, VGG,\\ ResNet, YOLO\end{tabular} & N.A. & N.A. & $\sim$0\% & N.A. \\ \hline
HAp-FT \cite{weiHApFTHybridApproximate2024} & \begin{tabular}[c]{@{}l@{}}Weight splitting,\\ approximate error checking\end{tabular} & \begin{tabular}[c]{@{}l@{}}Stuck-at, SEU /\\ RTL FI\end{tabular} & \begin{tabular}[c]{@{}l@{}}SCALE-Sim,\\ CACTI\end{tabular} & \begin{tabular}[c]{@{}l@{}}+2.7\%\\ (analytical)\end{tabular} & ResNet-20 & N.A. & +0.35\% & +0.15\% & N.A. \\ \cline{6-10} 
 &  &  &  &  & MobileNet-V2 & N.A. & +2.24\% & +0.06\% & N.A. \\ \cline{6-10} 
 &  &  &  &  & AlexNet & N.A. & +2.51\% & +0.18\% & N.A. \\ \hline
Salami et al. \cite{salamiResilienceRTLNN2018} & Word and bit masking & \begin{tabular}[c]{@{}l@{}}Stuck-at, SEU /\\ RTL FI\end{tabular} & \begin{tabular}[c]{@{}l@{}}HLS / FPGA\\ Xilinx VC707\end{tabular} & N.A. & Custom model & N.A. & N.A. & N.A. & N.A. \\ \hline
Li et al. \cite{liSoftErrorMitigation2020} & \begin{tabular}[c]{@{}l@{}}Periodic self-testing,\\ masking\end{tabular} & \begin{tabular}[c]{@{}l@{}}SEU /\\ Output flipping\end{tabular} & \begin{tabular}[c]{@{}l@{}}RTL / FPGA\\ Xilinx ZCU102\end{tabular} & \begin{tabular}[c]{@{}l@{}}\textless +17\% LUTs\\ \textless +9\% FFs\end{tabular} & \begin{tabular}[c]{@{}l@{}}MobileNet-V2,\\ ResNet-20,\\ GoogLeNet\end{tabular} & N.A. & N.A. & +0\% & N.A. \\ \hline
\multirow{2}{*}{Libano et al. \cite{libanoSelectiveHardeningNeural2019}} & \multirow{2}{*}{Selective layer TMR} & \multirow{2}{*}{\begin{tabular}[c]{@{}l@{}}SEU /\\ Neutron Beam, FI\end{tabular}} & \multirow{2}{*}{\begin{tabular}[c]{@{}l@{}}HLS / FPGA\\ Xilinx Zynq 7000\\ \& Ultrascale+\end{tabular}} & +6.7\% LUTs & MNIST CNN & N.A. & N.A. & N.A. & N.A. \\ \cline{5-10} 
 &  &  &  & \begin{tabular}[c]{@{}l@{}}+56.2\% LUTs\\ +5.7\% FFs\\ +200\% DSPs\end{tabular} & \begin{tabular}[c]{@{}l@{}}Iris Flower\\ ANN\end{tabular} & N.A. & N.A. & N.A. & N.A. \\ \hline
Syed et al.\textsuperscript{1} \cite{syedFPGAImplementationFaultTolerant2024} & \begin{tabular}[c]{@{}l@{}}Coarse reconfigurable\\ TMR\end{tabular} & \begin{tabular}[c]{@{}l@{}}SET, SEU, MBU /\\ RTL FI\end{tabular} & \begin{tabular}[c]{@{}l@{}}HLS / FPGA\\ Xilinx VCU118\end{tabular} & \textgreater{}200\% & \begin{tabular}[c]{@{}l@{}}Custom model\\ @200 MHz\end{tabular} & \begin{tabular}[c]{@{}l@{}}HP: 1866 mW\\ FT:  1911 mW\\ +2.4\%\end{tabular} & \begin{tabular}[c]{@{}l@{}}HP: 9.8µJ\\ FT: 29.9µJ\\ +204.9\%\end{tabular} & \begin{tabular}[c]{@{}l@{}}HP: 5.21 µs\\ FT: 15.7 µs\\ +197.6\%\end{tabular} & \begin{tabular}[c]{@{}l@{}}HP: 524.7 GOPs/W\\ FT: 171.1 GOPs/W\\ -67.4\%\end{tabular} \\ \hline
\begin{tabular}[c]{@{}l@{}}NEureka 4×4\textsuperscript{2} \cite{prasadSpecializationMeetsFlexibility2023}\end{tabular} & None & - & \begin{tabular}[c]{@{}l@{}}RTL / ASIC\\ GF 12nm\end{tabular} & \begin{tabular}[c]{@{}l@{}}0.235 mm²,\\ $\sim$1943 kGE\end{tabular} & \begin{tabular}[c]{@{}l@{}}Single Layer\textsuperscript{3}\\ @ 500 MHz,\\      0.80V\end{tabular} & 218.5 mW & 4.45 µJ & 20.36 µs & 2120.9 GOPs/W \\ \hline
\textbf{\begin{tabular}[c]{@{}l@{}}Safe-NEureka\\ 4×4\\ (this work)\end{tabular}} & \textbf{\begin{tabular}[c]{@{}l@{}}Fine grained reconfigurable\\ DMR (HMR),\\ ECC-protected memory interface,\\ TMR-protected controller\end{tabular}} & \textbf{\begin{tabular}[c]{@{}l@{}}SET, SEU /\\ Gate-level FI\end{tabular}} & \textbf{\begin{tabular}[c]{@{}l@{}}RTL / ASIC\\ GF 12nm\end{tabular}} & \textbf{\begin{tabular}[c]{@{}l@{}}0.271 mm²,\\ $\sim$2200 kGE /\\ +15\%\end{tabular}} & \textbf{\begin{tabular}[c]{@{}l@{}}Single Layer\textsuperscript{3}\\ @ 500 MHz,\\      0.80V\end{tabular}} & \textbf{\begin{tabular}[c]{@{}l@{}}PERF: 234.7 mW\\ RED: 241.7 mW\\ +3.0\%\end{tabular}} & \textbf{\begin{tabular}[c]{@{}l@{}}PERF: 5.0 µJ\\ RED: 9.4 µJ \\ +88.0\%\end{tabular}} & \textbf{\begin{tabular}[c]{@{}l@{}}PERF: 21.4 µs\\ RED: 38.94 µs\\ +82.0\%\end{tabular}} & \textbf{\begin{tabular}[c]{@{}l@{}}PERF: 1878.8 GOPs/W\\ RED: 1002.4 GOPs/W\\ -46.6\%\end{tabular}} \\ \cline{6-10} 
 &  &  &  &  & \textbf{\begin{tabular}[c]{@{}l@{}}MobileNet-V1\\ (res. 96$\times$96) \\ @ 500 MHz\end{tabular}} & \textbf{-} & \textbf{-} & \textbf{\begin{tabular}[c]{@{}l@{}}PERF: 2.94 ms\\ RED: 5.22 ms\\ +77.5\%\end{tabular}} & \textbf{-} \\ \cline{6-10} 
 &  &  &  &  & \textbf{\begin{tabular}[c]{@{}l@{}}ResNet-8\\ @ 500 MHz\end{tabular}} & \textbf{-} & \textbf{-} & \textbf{\begin{tabular}[c]{@{}l@{}}PERF: 0.94 ms\\ RED: 1.83 ms\\ +95\%\end{tabular}} & \textbf{-} \\ \hline
\end{tabular}%
}
\raggedright
\\ \footnotesize \textsuperscript{1} Aggregated power and energy for the three accelerators. Efficiency is computed on provided data. \hfill
\\ \footnotesize \textsuperscript{2} Figures obtained using the open-source design in the experimental setup of this work. \hfill
\\ \footnotesize \textsuperscript{3} Dense 3$\times$3 convolution $[k_{i},k_{o},h_{o},w_{o}]=[256,32,8,8]$ \hfill
\end{table}

FPGA-targeted designs often employ coarse-grained or model-specific strategies. The reconfigurable architecture proposed by \citet{syedFPGAImplementationFaultTolerant2024} shares the concept of dynamic redundancy. To avoid static costs, three accelerator instances can be grouped to perform the same task (\textit{FT} mode) or concurrent tasks (\textit{HP} mode). However, this design implements coarse-grained redundancy. As a result, error detection occurs only at the accelerator boundary, resulting in higher detection latency compared to fine-grained checking. To achieve correction, their approach necessitates \gls{tmr}, implying an area overhead exceeding 200\% (inherent to triplication). This cost is reflected in their operational metrics: when switching to \textit{FT} mode, latency and energy consumption increase by roughly 200\%, while efficiency drops by approximately 66\%. Furthermore, their implementation relies on hls4ml \cite{duarteFastInferenceDeep2018} and is tightly coupled to specialized branched or fused models, limiting flexibility.

Similarly, regarding \gls{hls}-based FPGA implementations, \citet{libanoSelectiveHardeningNeural2019} propose a selective layer-level hardening strategy that permanently commits resources based on offline vulnerability analysis. While effective for specific configurations, this approach relies on static heuristics and is deeply correlated to the specific model layer structure. Consequently, unlike dynamic datapath reconfiguration, it requires prior knowledge of the workload's sensitivity and cannot adapt to changing reliability requirements at runtime.

The Aphelios \gls{npu} \cite{sunApheliosSelectiveLockstep2025} shares our premise that full datapath replication is often prohibitively expensive. It employs \gls{dmr} on a downsized systolic array for selected neurons to reduce area costs. However, this redundancy is fixed, partial and non-reconfigurable. More critically, while Aphelios protects the datapath and applies \gls{dmr} to control logic, it lacks a mechanism for the controller to transparently recover from faults after detection. Implemented in Silvaco 15nm, it incurs overheads of 30.6\% in area and 33\% in energy.

Safe-NEureka establishes a fine-grained, fully deterministic \gls{rhbd} methodology rooted in the accelerator's microarchitecture. By partitioning the baseline 4$\times$4 datapath into two 4$\times$2 halves that can be dynamically configured for either parallel execution or redundant \gls{dmr} computation, our approach achieves the reliability of \gls{dmr} without the prohibitive cost of static replication. Unlike methods that rely on heuristics, partial \gls{pe} coverage, or specific data resilience properties, we guarantee bit-exact runtime recovery via hardware rollback, protecting the full datapath independent of the target \gls{ai} model. Furthermore, we achieve accelerator-wide protection by complementing \gls{hmr} with \gls{ecc}-protected memory interfaces and a \gls{tmr}-hardened controller. The power, performance, and area of the proposed methodology are exhaustively validated via a GlobalFoundries 12nm tapeout-ready implementation, alongside fault coverage analysis on synthesized netlists.
\section{Conclusion}

This work presented Safe-NEureka, a run-time reconfigurable, fault-tolerant \gls{dnn} accelerator optimized for heterogeneous RISC-V clusters. To address the varying requirements of space-based AI, the architecture features a hybrid datapath that can be split into two independent halves. This flexibility allows the system to dynamically toggle between a high-throughput \textit{performance} mode and a reliable \textit{redundancy} mode. In the latter, fault tolerance is ensured via a transparent \gls{dmr} strategy with low-overhead hardware recovery, complemented by \gls{ecc} on memory and \gls{tmr} on the controller.
Implemented in GlobalFoundries 12nm technology, the architecture incurs a modest 15\% area overhead compared to a non-fault-tolerant baseline. In performance mode, the design maintains near-native speeds with only a 5\% throughput penalty and an 11\% reduction in energy efficiency. Conversely, while redundancy mode reduces throughput by 48\% and efficiency by 53\% due to replication costs, it achieves a 96\% reduction in incorrect results. Ultimately, this flexibility ensures that the accelerator can be dynamically tuned to the criticality of the task at runtime, satisfying mixed-criticality constraints.


\bibliographystyle{ACM-Reference-Format}
\bibliography{bibliography/zotero}

\end{document}